\documentclass[preprint,authoryear,review,12pt]{elsarticle}

\usepackage{amssymb}
\usepackage{amsthm}
\usepackage{amsmath}
\usepackage{latexsym}

\usepackage{pstricks}
\usepackage{epsfig}

\usepackage{graphicx}
\usepackage{caption}
\usepackage{subcaption}

\usepackage{url}

\usepackage[utf8]{inputenc}

\newtheorem{definition}{Definition}

\newtheorem{lema}{Lemma}
\newtheorem{theorem}{Theorem}

\begin{document}


\begin{center}
\Large Maximal skewness projections for scale mixtures of skew-normal vectors
\end{center}

\bigskip

\bigskip

\bigskip

\bigskip

\begin{center}

{\bf Jorge M Arevalillo, PhD}

Data Scientist, University Nacional Educación a Distancia, Dpt. Statistics and Operational Research.
Paseo Senda del Rey 9 28040 Madrid, Spain

Tel: +34 91 398 72 64
Email: jmartin@ccia.uned.es
\end{center}

\begin{center}

{\bf Hilario Navarro, PhD}

University Nacional Educación a Distancia, Dpt. Statistics and Operational Research.
Paseo Senda del Rey 9 28040 Madrid, Spain

Tel: +34 91 398 72 55
Email: hnavarro@ccia.uned.es
\end{center}

\newpage

{\Large\bf Abstract}

\bigskip

\bigskip

Multivariate scale mixtures of skew-normal (SMSN) variables are flexible models that account for non-normality in multivariate data scenarios by tail weight assessment and a shape vector representing the asymmetry of the model in a directional fashion. Its stochastic representation involves a skew-normal (SN) vector and a non negative mixing scalar variable, independent of the SN vector, that injects kurtosis into the SMSN model. We address the problem of finding the maximal skewness projection for vectors that follow a SMSN distribution; when simple conditions on the moments of the mixing variable are fulfilled, it can be shown that the direction yielding the maximal skewness is proportional to the shape vector. This finding stresses the directional nature of the asymmetry in this class of distributions; it also provides the theoretical foundations for solving the skewness model based projection pursuit for SMSN vectors. Some examples that show the validity of our theoretical findings for the most famous distributions within the SMSN family are also given. For the sake of completeness we carry out a simulation experiment with artificial data, which sheds light on the usefulness and implications of our result in the statistical practice.

\bigskip

\bigskip

 Scale mixture of Skew-normal distributions. Maximal skewness. Moments. Mixing variable.

\newpage




\title{On the direction maximizing non-normality for scale mixtures of skew-normal vectors}


\author{Jorge M Arevalillo\corref{cor1}}
\ead{jmartin@ccia.uned.es}

\author{Hilario Navarro}
\ead{hnavarro@ccia.uned.es}

\cortext[cor1]{Corresponding author}

\address{Department of Statistics, Operational Research and
Numerical Analysis, University Nacional Educación a Distancia (UNED)
Paseo Senda del Rey 9, 28040, Madrid, Spain}

\section{Introduction}
\label{Intro}

The multivariate skew-normal (SN) distribution is a flexible model widely accepted to regulate asymmetry departures from normality. The study of its theoretical properties and applications has originated vast research \citep{AzzaliniCapitanio1999,CapitanioAzzaliniStanghellini,Azzalini2005,Contreras2012,BalakrishnanScarpa,BalakrishnanCapitanioScarpa}. We adopt the notation of the seminal works by \cite{AZZALINI1996} and \cite{AzzaliniCapitanio1999} to define the density function of a $p$-dimensional SN vector with location vector $\boldsymbol\xi=(\xi_1,\ldots,\xi_p)^\prime$ and scale matrix $\boldsymbol\Omega$ as follows:

\begin{equation}
f(\boldsymbol x; \boldsymbol\xi,\boldsymbol\alpha,\boldsymbol\Omega)=2\phi_p(\boldsymbol x-\boldsymbol\xi;\boldsymbol\Omega)\Phi(\boldsymbol\alpha^\prime\boldsymbol\omega^{-1}(\boldsymbol x-\boldsymbol\xi))\hspace{0.25cm}:\hspace{0.25cm}\boldsymbol{x}\in\mathbb{R}^p,
\label{EqSN}
\end{equation}

\noindent where $\phi_p(\cdot;\boldsymbol\Omega)$ denotes the $p$-dimensional normal density with zero mean and covariance matrix $\boldsymbol\Omega$, $\Phi$ is the distribution function of a standard $N(0,1)$ variable, $\boldsymbol\omega=diag(\omega_1,\ldots,\omega_p)$ is a scale diagonal matrix with non negative entries such that $\bar{\boldsymbol\Omega}=\boldsymbol\omega^{-1}\boldsymbol\Omega\boldsymbol\omega^{-1}$ is a correlation matrix and $\boldsymbol\alpha$ is a $p$-dimensional shape parameter that regulates the skewness. Note that the scale matrix $\boldsymbol\omega$ can be written as $\boldsymbol\omega=(\boldsymbol\Omega\odot\boldsymbol I_p)^{1/2}$, where the symbol $\odot$ denotes the entry-wise matrix product.

We will write $\boldsymbol X \sim SN_p(\boldsymbol\xi,\boldsymbol\Omega,\boldsymbol\alpha)$ to denote that $\boldsymbol X$ follows a $p$-dimensional skew-normal distribution with density function (\ref{EqSN}), with $\boldsymbol X\sim N_p(\boldsymbol\xi,\boldsymbol\Omega)$ when $\boldsymbol\alpha = \boldsymbol 0$. We can also observe that $\boldsymbol X=\boldsymbol\xi+\boldsymbol\omega\boldsymbol Z$, where $\boldsymbol Z$ is a normalized multivariate skew-normal variable with density function given by

\begin{equation}
f(\boldsymbol z; \boldsymbol 0,\boldsymbol\alpha,\boldsymbol\Omega)=2\phi_p(\boldsymbol z;\bar{\boldsymbol\Omega})\Phi(\boldsymbol\alpha^\prime\boldsymbol z).
\label{EqNormalizedSN}
\end{equation}

The multivariate scale mixture of skew-normal (SMSN) distribution is an extension of the SN model that incorporates an additional parameter to handle kurtosis departures from normality \citep{BrancoDey2001}. The SMSN family has become increasingly popular because it defines a wide class of distributions for handling skewness and kurtosis simultaneously; the family contains some popular multivariate models, like the skew-t or the double exponential.

This paper explores the projection pursuit problem when the underlying multivariate model belongs to the class of SMSN distributions. Specifically, when handling non normal data, one may be interested in finding ``relevant'' projections as those ones maximizing a nonnormality measure \citep{Huber1985}. The problem was addressed for SN vectors by \cite{Loperfido2010}, who also suggested its extension to a more general framework. In this paper we revisit the problem and extend it by exploring the projections that maximize skewness for vectors that follow a multivariate SMSN distribution. Conditions on the moments of the mixing variable, that allow to find an analytical solution to the problem, are studied and the role of the shape vector that parameterizes the asymmetry of the model is discussed; several examples that illustrate the results of the theory are also given in order to highlight the theoretical insights. The rest of the manuscript is organized as follows: the next section gives a brief introduction about SMSN distributions. In Section \ref{MaxSkewDirection} we address the problem of finding the maximal skewness projection for SMSN vectors; some examples that shed light on the theory are presented in Section \ref{Examples}. In Section \ref{Numerical_work} the theoretical findings are illustrated through a simulation experiment. The paper is finished giving some concluding remarks.

\section{Scale mixtures of skew-normal distributions}
\label{ScaleMixturesSN}

The family of multivariate SMSN variables was introduced by \cite{BrancoDey2001} as a subclass of the more general class of skew-elliptical distributions. The class of SMSN distributions is essentially characterized by the product of a SN vector and an independent non negative scalar variable; the former controls the non-normality of the multivariate distribution described in terms of asymmetry while the later injects kurtosis in the resulting multivariate model. Some deeper insights about the properties and features of SMSN variables came up with \cite{Capitanio:arXiv1207.0797}, who extended some previous work on the canonical transformation of SN vector to the family of SMSN variables.

In this paper we use the notation adopted by \cite{Capitanio:arXiv1207.0797} to define the multivariate SMSN distributions as follows.

\begin{definition}
Let $\boldsymbol Z$ be a random vector such that $\boldsymbol Z \sim SN_p(\boldsymbol 0,\bar{\boldsymbol\Omega},\boldsymbol\alpha)$, with density function (\ref{EqNormalizedSN}), and let $S$ be a non negative scalar variable, independent of $\boldsymbol Z$. The random vector $\boldsymbol X=\boldsymbol\xi+\boldsymbol\omega S\boldsymbol Z$, where $\boldsymbol\omega$ is a scale diagonal matrix, is said to follow a multivariate SMSN distribution.
\label{DefSMSN}
\end{definition}

We can scale the correlation matrix $\bar{\boldsymbol\Omega}$ to obtain the full rank scale matrix given by $\boldsymbol\Omega=\boldsymbol\omega\bar{\boldsymbol\Omega}\boldsymbol\omega$, so we write $\boldsymbol X \sim SMSN_p(\boldsymbol\xi,\boldsymbol\Omega,\boldsymbol\alpha,H)$, with $H$ denoting the univariate distribution function of the mixing variable $S$, to indicate that $\boldsymbol X$ follows the multivariate SMSN distribution. Note that if we take $\boldsymbol\alpha\boldsymbol =\boldsymbol 0$ then $\boldsymbol X$ becomes a scale mixture of multivariate normal distributions, a subclass of the elliptically contoured multivariate distributions. Note also that, when $H$ is a degenerate distribution at $S=1$ we have $\boldsymbol X \sim SN_p(\boldsymbol\xi,\boldsymbol\Omega,\boldsymbol\alpha)$.

\section{Skewness maximization}
\label{MaxSkewDirection}

Let us consider a vector $\boldsymbol X$ such that $\boldsymbol X \sim SMSN_p(\boldsymbol\xi,\boldsymbol\Omega,\boldsymbol\alpha,H)$ and the scaled vector $\boldsymbol U=\boldsymbol\Sigma^{-1/2}(\boldsymbol X -\boldsymbol\xi)$ with $\boldsymbol\Sigma$ the covariance matrix of $\boldsymbol X$. In accordance to Definition \ref{DefSMSN}, the vector admits the following stochastic representation: $\boldsymbol X=\boldsymbol\xi+\boldsymbol\omega S\boldsymbol Z=\boldsymbol\xi+S\boldsymbol Z^*$, with $\boldsymbol Z^* \sim SN_p(0,\boldsymbol\Omega,\boldsymbol\eta)$ where $\boldsymbol\eta=\boldsymbol\omega^{-1}\boldsymbol\alpha$  and $\boldsymbol\Omega=\boldsymbol\omega\bar{\boldsymbol\Omega}\boldsymbol\omega$.

We address the problem of finding the direction $\boldsymbol c$ for which the scalar variable $Y=\boldsymbol {c^\prime  U}$ attains the maximum skewness. Thus, our goal is to solve the following optimization problem: $\displaystyle\max_{\boldsymbol c\in\mathbb{R}_0^p}\gamma_1(\boldsymbol {c^\prime  U})$, with $\gamma_1$ the skewness index defined by $\displaystyle\gamma_1(Y)=E^2\left(\frac{Y-\mu_Y}{\sigma_Y}\right)^3$ and $\mathbb{R}_0^p$ the set of all non-null $p$-dimensional vectors.

Since $\gamma_1$ is scale invariant, we can confine to vectors such that $\boldsymbol c^\prime\boldsymbol c=1$; hence, the problem of finding the directional skewness can be described as

\begin{equation}
\gamma_{1,p}^D(\boldsymbol X)=\max_{\boldsymbol c\in\mathbb{S}_p}\gamma_1(\boldsymbol {c^\prime U})
\label{EqSkewnessMaximization}
\end{equation}

\noindent where $\mathbb{S}_p=\{\boldsymbol c\in {\mathbb{R}^p} : \boldsymbol c^\prime\boldsymbol c=1\}$.

Alternatively, it can be stated by the following equivalent formulation:

\begin{equation}
\gamma_{1,p}^D(\boldsymbol X)=\max_{\boldsymbol d\in\mathbb{S}_p^*}\gamma_1(\boldsymbol {d^\prime X})
\label{EqSkewnessMaximization_bis}
\end{equation}
\noindent where $\boldsymbol d =\boldsymbol\Sigma^{-1/2}\boldsymbol c$ with $\mathbb{S}_p^*=\{\boldsymbol d\in {\mathbb{R}^p} : \boldsymbol d^\prime\boldsymbol\Sigma\boldsymbol d=1\}$.

The solutions of any of the previous equivalent problems are given by

\begin{equation}
\boldsymbol\lambda_{\boldsymbol X} = \mbox{arg}\max_{\boldsymbol d\in\mathbb{S}_p^*}\gamma_1(\boldsymbol {d^\prime X})\mbox{   ,   }
\boldsymbol\lambda_{\boldsymbol U} = \mbox{arg}\max_{\boldsymbol c\in\mathbb{S}_p}\gamma_1(\boldsymbol {c^\prime U})
\label{Eq_Sol_arg}
\end{equation}

\noindent both satisfying that $\displaystyle \boldsymbol\lambda_{\boldsymbol X} \propto\boldsymbol\Sigma^{-1/2}\boldsymbol\lambda_{\boldsymbol U}$.

\subsection{Main contribution}

Before proving the main contribution of the paper, we need the following auxiliary lemma.

\begin{lema}
\label{AuxiliaryLemaSigma} Let $\boldsymbol X$ be a random vector such that $\boldsymbol X \sim SMSN_p(\boldsymbol \xi,\boldsymbol\Omega,\boldsymbol\alpha,H)$. Let us assume that the mixing variable $S$ has finite second order moment. If  $\boldsymbol\Sigma$ is the covariance matrix of $\boldsymbol X$ then $\boldsymbol\Sigma^{-1}\boldsymbol\gamma$, where $\displaystyle\boldsymbol\gamma=\frac{\boldsymbol\Omega\boldsymbol\eta}{\sqrt{1+\boldsymbol\eta^\prime\boldsymbol\Omega\boldsymbol\eta}}$, is proportional to $\boldsymbol\eta=\boldsymbol\omega^{-1}\boldsymbol\alpha$.
\end{lema}

{\noindent\bf Proof.} The covariance matrix for SMSN vectors is given by

$$\displaystyle\boldsymbol\Sigma=c_2 \boldsymbol\Omega -\frac{2}{\pi} c_1^2 \boldsymbol\gamma\boldsymbol\gamma^\prime =
c_2\left(\boldsymbol\Omega-\frac{2}{\pi}\frac{c_1^2}{c_2}\boldsymbol\gamma\boldsymbol\gamma^\prime\right),$$

\noindent with $c_1=E(S)$ and $c_2=E(S^2)$ \citep{Capitanio:arXiv1207.0797,AzzaliniCapitanio2014}.

In order to calculate $\boldsymbol\Sigma^{-1}$, we use the well-known Sherman-Morrison formula, which is given by

\begin{equation}
\boldsymbol (\boldsymbol A+\boldsymbol u \boldsymbol v^\prime)^{-1} = \boldsymbol A^{-1}-\frac{\boldsymbol A^{-1} \boldsymbol u \boldsymbol v^\prime \boldsymbol A^{-1}}{1+\boldsymbol v^\prime \boldsymbol A^{-1} \boldsymbol u}.
\label{ShermanMorrisonWoodbury}
\end{equation}

Taking $A=\boldsymbol\Omega$, $\displaystyle \boldsymbol u= -\frac{2}{\pi}\frac{c_1^2}{c_2} \boldsymbol\gamma$, and $\boldsymbol v=\boldsymbol\gamma$ we obtain that

$$c_2\boldsymbol\Sigma^{-1}=\left(\boldsymbol\Omega^{-1}-\frac{\boldsymbol\Omega^{-1}\boldsymbol
\gamma\left(-\frac{2}{\pi}\frac{c_1^2}{c_2}\right)\boldsymbol\gamma^\prime\boldsymbol\Omega^{-1}}{1+\left(-\frac{2}{\pi}\frac{c_1^2}{c_2}\right)\boldsymbol\gamma^\prime\boldsymbol\Omega^{-1}\boldsymbol\gamma}\right)
=\left(\boldsymbol\Omega^{-1}+\frac{\boldsymbol\Omega^{-1}\boldsymbol
\gamma\boldsymbol\gamma^\prime\boldsymbol\Omega^{-1}}{\frac{\pi}{2}\frac{c_2}{c_1^2}-\boldsymbol\gamma^\prime\boldsymbol\Omega^{-1}\boldsymbol\gamma}\right)$$


\noindent from which we get $\displaystyle\boldsymbol\Sigma^{-1}\boldsymbol\gamma=
\frac{\boldsymbol\Omega^{-1}\boldsymbol\gamma}{c_2-\frac{2}{\pi}c_1^2\boldsymbol\gamma^\prime\boldsymbol\Omega^{-1}\boldsymbol\gamma}$ after some calculations. Consequently, $\displaystyle\boldsymbol\Sigma^{-1}\boldsymbol\gamma$ is proportional to
$\displaystyle\boldsymbol\Omega^{-1}\boldsymbol\gamma=\frac{\boldsymbol\eta}{\sqrt{1+\boldsymbol\eta^\prime\boldsymbol\Omega\boldsymbol\eta}}$, which implies the assertion of the statement $\blacksquare$

The quantity $\gamma_1$ in (\ref{EqSkewnessMaximization}) is a multivariate skewness index that captures the directional nature of the asymmetry \citep{MalkovichAfifi1973}. Although it depends on the form of the stochastic representation of the SMSN vector, specifically on the distribution of the mixing variable $S$, we can show that the vector yielding the maximum skewness lies on the direction of the shape parameter $\boldsymbol\eta$. This happens when it holds a rather simple condition on the moments of the mixing variable, as shown by the next theorem.

\begin{theorem}
Let $\boldsymbol X$ be a random vector such that $\boldsymbol X \sim SMSN_p(\boldsymbol \xi,\boldsymbol\Omega,\boldsymbol\alpha,H)$ such that the moment inequality for the mixing variable: $\displaystyle\frac{4}{\pi}E^2(S)\geq E(S^2)$ holds. Then the maximum skewness in (\ref{EqSkewnessMaximization_bis}) is attained at the direction of the vector $\boldsymbol\eta^\prime=\boldsymbol\alpha^\prime\boldsymbol\omega^{-1}$.
\label{TheMaxProjectSkew}
\end{theorem}

{\noindent\bf Proof.} Taking into account the following equivalent restrictions: $\boldsymbol c \in \mathbb{S}_p$ or $\boldsymbol d \in \mathbb{S}_p^*$ from (\ref{EqSkewnessMaximization}) or (\ref{EqSkewnessMaximization_bis}), we get

$$\gamma_1(Y)=\gamma_1(\boldsymbol c^\prime \boldsymbol U)=E^2[\boldsymbol c^\prime (\boldsymbol U-E(\boldsymbol U))]^3=E^2\left[\boldsymbol d^\prime\left(\boldsymbol X-\boldsymbol\xi-E(S)\sqrt{\frac{2}{\pi}}\boldsymbol\gamma\right)\right]^3,$$

\noindent with $\boldsymbol d=\boldsymbol\Sigma^{-1/2}\boldsymbol c$ and
$\displaystyle\boldsymbol\gamma=\boldsymbol\omega\boldsymbol\delta=\frac{\boldsymbol\omega\bar{\boldsymbol\Omega}\boldsymbol\alpha}{\sqrt{1+\boldsymbol\alpha^\prime\bar{\boldsymbol\Omega}\boldsymbol\alpha}}=
\frac{\boldsymbol\Omega\boldsymbol\eta}{\sqrt{1+\boldsymbol\eta^\prime\boldsymbol\Omega\boldsymbol\eta}}$.






The previous expression for $\gamma_1$ admits the following reformulations:

\begin{equation}
\gamma_1(Y)=\gamma_1(\boldsymbol c^\prime \boldsymbol U)=\gamma_1(\boldsymbol d^\prime \boldsymbol X)=\gamma_1(\boldsymbol d^\prime (\boldsymbol X-\boldsymbol\xi))
\label{Equations_gamma1}
\end{equation}

\noindent where $\displaystyle \boldsymbol d^\prime (\boldsymbol X-\boldsymbol\xi)=SZ_0$ and $Z_0=\boldsymbol d^\prime\boldsymbol Z^*$ is a scalar random variable such that $Z_0\sim SN_1\left(0,\omega_{\boldsymbol d},\alpha_{\boldsymbol d}\right)$ where
$\displaystyle\alpha_{\boldsymbol d}=\frac{\boldsymbol d^\prime\boldsymbol\gamma}{\sqrt{\omega_{\boldsymbol d}-(\boldsymbol d^\prime\boldsymbol\gamma)^2}}
=\frac{\sqrt{t}}{\sqrt{\omega_{\boldsymbol d}-t}}$ and $\displaystyle\omega_{\boldsymbol d}=\boldsymbol d^\prime\boldsymbol\Omega\boldsymbol d$ with $t=(\boldsymbol d^\prime\boldsymbol\gamma)^2$ \cite[formula (5.44)]{AzzaliniCapitanio2014}. Alternatively, $Z_0$ can be represented by $Z_0=\omega_{\boldsymbol d}^{1/2}U_0$, where $U_0\sim SN_1\left(0,1,\alpha_{\boldsymbol d}\right)$.

In order to find an analytical expression for the quantity in (\ref{Equations_gamma1}), we need the moments of $U_0$ up to the third one. They are given by

$$E(U_0)=\sqrt{\frac{2}{\pi}}\delta_0\mbox{  ,  }E(U_0^2)=1\mbox{  and  }E(U_0^3)=\sqrt{\frac{2}{\pi}}(3 \delta_0-\delta_0^3),$$

\noindent where $\displaystyle\delta_0^2=\frac{\alpha_{\boldsymbol d}^2}{1+\alpha_{\boldsymbol d}^2}=\omega_{\boldsymbol d}^{-1}(\boldsymbol d^\prime\boldsymbol\gamma)^2=\omega_{\boldsymbol d}^{-1}t$ is a quantity that satisfies the following inequality: $0 \leq \displaystyle\delta_0^2=\omega_{\boldsymbol d}^{-1}t \leq 1$. Inserting the moments of the variable $U_0$ into (\ref{Equations_gamma1}), we get

$$\gamma_1(\boldsymbol d^\prime (\boldsymbol X-\boldsymbol\xi))=E^2[(SZ_0-E(SZ_0))^3]=\omega_{\boldsymbol d}^3E^2[(SU_0-E(SU_0))^3]$$

$$=\omega_{\boldsymbol d}^3[E(S^3)E(U_0^3)-3E(S^2)E(S)E(U_0^2)E(U_0)+2E^3(S)E^3(U_0)]^2$$

$$=\omega_{\boldsymbol d}^3
\left[E(S^3)\sqrt{\frac{2}{\pi}}(3\delta_0-\delta_0^3)-3E(S^2)E(S)\sqrt{\frac{2}{\pi}}\delta_0+2E^3(S)\sqrt{\frac{2}{\pi}}\frac{2}{\pi}\delta_0^3\right]^2$$

$$=\frac{2}{\pi}\omega_{\boldsymbol d}^3\delta_0^2[a\delta_0^2-3b]^2=\frac{2}{\pi}t[at-3b\omega_{\boldsymbol d}]^2,$$

\noindent where $\displaystyle a=\frac{4}{\pi}E^3(S)-E(S^3)$ and $\displaystyle b=E(S)E(S^2)-E(S^3)$ are terms that depend on the moments of the mixing variable.


From the previous arguments it can be shown that $\gamma_1(Y)=\gamma_1(\boldsymbol c^\prime \boldsymbol U)=\gamma_1(\boldsymbol d^\prime (\boldsymbol X-\boldsymbol\xi)$ is a function of the quantity $t=(\boldsymbol d^\prime\boldsymbol\gamma)^2$, specifically

\begin{equation}
\gamma_1(Y)=h(t)=\frac{2}{\pi}t[at-3b\omega_{\boldsymbol d}]^2
\label{Function_Ht}
\end{equation}

\noindent where as before $\displaystyle\boldsymbol\gamma=\boldsymbol\omega\boldsymbol\delta=\frac{\boldsymbol\Omega\boldsymbol\eta}{\sqrt{1+\boldsymbol\eta^\prime\boldsymbol\Omega\boldsymbol\eta}}$ and the terms $a$ and $b$ are given by the previous moment expressions.

Firstly, we prove the non decreasing behaviour of $h(t)$, whose first derivative is given by

$$\displaystyle h^\prime(t)=\frac{6\omega_{\boldsymbol d}^2}{\pi}\left(\frac{at}{\omega_{\boldsymbol d}}-3b\right)
\left(\frac{at}{\omega_{\boldsymbol d}}-b-\frac{2bct}{\omega_{\boldsymbol d}}\right)\mbox{  :  }0 \leq\omega_{\boldsymbol d}^{-1}t\leq 1$$

\noindent with $a$ and $b$ as previously defined and the quantity $c$ given by $\displaystyle c=\frac{2}{\pi}\frac{E^2(S)}{E(S^2)}$.

The well-known moment inequality $E(S^3)\geq E(S)E(S^2)$ implies that $b \leq 0$, so we are going to distinguish two cases: if $a \geq 0$ then $h^\prime(t)> 0$ and $h(t)$ is a non decreasing function. On the other hand, when $a<0$ the condition on the moments of $S$ from the statement implies that $b\leq a$, which in turn gives $\displaystyle\frac{b}{a}\geq 1$. Taking into account that $\displaystyle 0\leq\omega_{\boldsymbol d}^{-1}t\leq 1$ we can assert that
$\displaystyle\frac{at}{\omega_{\boldsymbol d}}-b>0$ from which we obtain that $h^\prime (t)>0$ and once again we conclude that $h(t)$ is a non decreasing function.

Taking into account that $h(t)$ is non decreasing, its maximum is attained at the maximum value of $\displaystyle t=(\boldsymbol d^\prime\boldsymbol\gamma)^2$; so our problem in (\ref{Eq_Sol_arg}) can be reduced to finding the direction that maximizes $(\boldsymbol d^\prime\boldsymbol\gamma)^2$. We know that

$$(\boldsymbol d^\prime\boldsymbol\gamma)^2=(\boldsymbol c^\prime\boldsymbol\Sigma^{-1/2}\boldsymbol\gamma)^2\leq (\boldsymbol c^\prime\boldsymbol c)
(\boldsymbol\Sigma^{-1/2}\boldsymbol\gamma)^\prime(\boldsymbol\Sigma^{-1/2}\boldsymbol\gamma)=\boldsymbol\gamma^\prime\boldsymbol\Sigma^{-1}\boldsymbol\gamma$$

\noindent and the maximum of the scalar product is attained when vector $\boldsymbol c$ is proportional to $\displaystyle\boldsymbol\Sigma^{-1/2}\boldsymbol\gamma$.

Hence, we can conclude that $\displaystyle\boldsymbol\lambda_U \propto\boldsymbol\Sigma^{-1/2}\boldsymbol\gamma$ which, taking into account that $\displaystyle \boldsymbol\lambda_X \propto\boldsymbol\Sigma^{-1/2}\boldsymbol\lambda_U$ from (\ref{Eq_Sol_arg}), implies that $\displaystyle\boldsymbol\lambda_X \propto\boldsymbol\Sigma^{-1}\boldsymbol\gamma$. This finding, together with the result of Lemma \ref{AuxiliaryLemaSigma}, proves the statement  $\blacksquare$

It is worthwhile noting that the condition $\displaystyle a=\frac{4}{\pi}E^3(S)-E(S^3)\geq 0$ ensures the validity of Theorem \ref{TheMaxProjectSkew}; in fact, it implies the condition $\displaystyle\frac{4}{\pi}E^2(S)-E(S^2)\geq 0$ because $b=E(S)E(S^2)-E(S^3)\leq 0$. However, the condition on the third moment of $S$ is not always met for some well-known subfamilies within the class of SNSM distributions. Hence, as we have discussed in the proof, the condition $\displaystyle \frac{4}{\pi}E^2(S)\geq E(S^2)$ should be used in the cases where $a\leq 0$. In the following examples we elaborate on this issues for particular cases where the moments of the mixing variable can be calculated explicitly.

\subsection{Examples}
\label{Examples}

In this section we present particular forms of the stochastic representation of the SMSN vector for which the result of Theorem \ref{TheMaxProjectSkew} is valid.

\subsubsection{The multivariate SN distribution}
\label{Ex.SN}

The SN multivariate model is obtained when the mixing variable of the $SMSN$ vector is degenerate at $S=1$. In this case $a\geq 0$ and the result of Theorem \ref{TheMaxProjectSkew} is verified. This finding was previously achieved by \cite{Loperfido2010} using a slight different approach.

\subsubsection{The multivariate skew-t distribution}
\label{Ex.Skew.t}

The multivariate ST distribution arises when the mixing variable of the SNSM vector is $S=V^{-1/2}$ with $V\sim\chi_\nu^2/\nu$. In this case, as stated by \cite{AzzaliniCapitanio2003}, we obtain that the density function of $\boldsymbol X$ is given by

\begin{equation}
f(\boldsymbol x; \boldsymbol\xi,\boldsymbol\Omega,\boldsymbol\alpha ,\nu)=2\,t_p(\boldsymbol x;\nu)T_1\left(\boldsymbol\alpha^\prime\boldsymbol\omega^{-1}(\boldsymbol x-\boldsymbol\xi)\left(\frac{\nu+p}{Q_{\boldsymbol x}+\nu}\right)^{1/2}; \nu+p\right)\mbox{ : }\boldsymbol{x}\in\mathbb{R}^p
\label{EqDensityST}
\end{equation}

\begin{figure}[h]
        \centering
        \begin{subfigure}[b]{0.5\textwidth}
                \includegraphics[height=3.5cm,width=7cm]{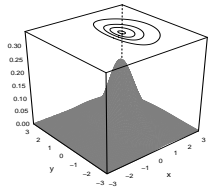}
                \caption{$\boldsymbol\alpha^\prime = (3,0)$}
                \label{Dens_skewt_3.0}
        \end{subfigure}%
        \begin{subfigure}[b]{0.5\textwidth}
                \includegraphics[height=3.5cm,width=7cm]{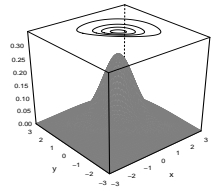}
                \caption{$\boldsymbol\alpha^\prime = (3,3)$}
                \label{Dens_skewt_3.3}
        \end{subfigure}
        \\
        \begin{subfigure}[b]{0.5\textwidth}
                \includegraphics[height=3.5cm,width=7cm]{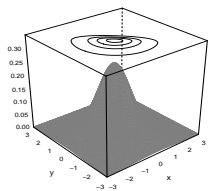}
                \caption{$\boldsymbol\alpha^\prime = (-3,-3)$}
                \label{Dens_skewt_n3.n3}
        \end{subfigure}%
        \begin{subfigure}[b]{0.5\textwidth}
                \includegraphics[height=3.5cm,width=7cm]{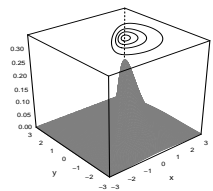}
                \caption{$\boldsymbol\alpha^\prime = (3,-3)$}
                \label{Dens_skewt_3.n3}
        \end{subfigure}
        \caption{Density functions of the bivariate ST variable, with location $\boldsymbol\xi = (0,0)$, scale matrix $\boldsymbol\Omega = \boldsymbol I_2$ and $\nu=4$, for different shape vectors.}
        \label{Fig:SkewT}
\end{figure}

We write $\boldsymbol X \sim ST_p(\boldsymbol\xi,\boldsymbol\Omega,\boldsymbol\alpha,\nu)$ to denote that $\boldsymbol X$ follows a $p$-dimensional ST distribution with density function (\ref{EqDensityST}). Figure \ref{Fig:SkewT} shows how the vector $\boldsymbol\alpha$ deforms the symmetry of the $t$ distribution when the asymmetry is injected through different directions; the contoured plots for each density function are depicted as well. It is worthwhile noting that when $\nu\rightarrow\infty$ the multivariate ST becomes a $p$-dimensional SN distribution, i.e. $\boldsymbol X \sim SN_p(\boldsymbol\xi,\boldsymbol\Omega,\boldsymbol\alpha)$.

Recall that the moments of the mixing variable are given by $\displaystyle E(S^k)=\frac{(\nu/2)^{k/2}\Gamma\left(\frac{\nu -k}{2}\right)}{\Gamma\left(\frac{\nu}{2}\right)}$ provided that $\nu >k : k\geq 1$. From this expression we obtain that for $\nu >3$

$$a=E(S)\left(\frac{4}{\pi}E^2(S)-\frac{\nu}{\nu-3}\right)=\nu E(S)\left(\frac{2}{\pi}\frac{\Gamma^2\left(\frac{\nu-1}{2}\right)}{\Gamma^2\left(\frac{\nu}{2}\right)}-\frac{1}{\nu-3}\right),$$

\noindent which is negative when $\nu < 9$ and positive when $\nu\geq 9$.

Using Lemma 1 from \cite{ArevalilloNavarro2015} we can easily state the validity of the condition $\displaystyle\frac{4}{\pi}E^2(S)\geq E(S^2)$.

\subsubsection{The multivariate skew double exponential distribution}
\label{Ex.Double.Exp}

The multivariate double exponential (DE) distribution was introduced as a generalization of its univariate counterpart. We say that $\boldsymbol X$ follows a $p$-dimensional multivariate DE distribution with location vector $\boldsymbol\xi$ and full rank scale matrix $\boldsymbol\Omega$ if its density function is given by

\begin{equation}
f(\boldsymbol x; \boldsymbol\xi,\boldsymbol\Omega)=\frac{\Gamma(\frac{p}{2})}{\pi^{p/2}\Gamma(p)2^{1+p}}|\boldsymbol\Omega|^{-1/2}
\exp\left\{-\frac{1}{2}\left[(\boldsymbol x-\boldsymbol\xi)^\prime\boldsymbol\Omega^{-1}(\boldsymbol x-\boldsymbol\xi)\right]^{1/2}\right\},
\label{EqMultPOTEX}
\end{equation}

The multivariate double exponential distribution can be seen as a scale mixture of multivariate normal variables with mixing variate $S=W^{1/2}$, where $W$ is a $\displaystyle Gamma\left(\frac{p+1}{2},\frac{1}{8}\right)$ \citep{Gomez2006}. When we take $W^{1/2}$ as the mixing variable in Definition \ref{DefSMSN} we get the SMSN variables defining the multivariate skew double exponential (SDE) distribution. We write $\boldsymbol X \sim SDE_p(\boldsymbol\xi,\boldsymbol\Omega,\boldsymbol\alpha)$ to indicate that $\boldsymbol X$ follows a $p$-dimensional SDE distribution with location $\boldsymbol\xi$, scale matrix $\boldsymbol\Omega=\boldsymbol\omega\bar{\boldsymbol\Omega}\boldsymbol\omega$ and shape asymmetry vector $\boldsymbol\alpha$.

\begin{figure}[h]
        \centering
        \begin{subfigure}[b]{0.5\textwidth}
                \includegraphics[height=3.5cm,width=7cm]{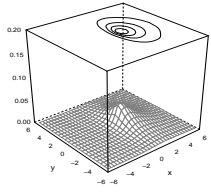}
                \caption{$\boldsymbol\alpha^\prime = (3,0)$}
                \label{Dens_SkewDoubleExp_3.0}
        \end{subfigure}%
        \begin{subfigure}[b]{0.5\textwidth}
                \includegraphics[height=3.5cm,width=7cm]{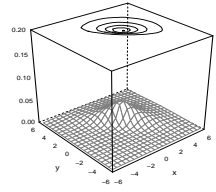}
                \caption{$\boldsymbol\alpha^\prime = (3,3)$}
                \label{Dens_SkewDoubleExp_3.3}
        \end{subfigure}
        \\
        \begin{subfigure}[b]{0.5\textwidth}
                \includegraphics[height=3.5cm,width=7cm]{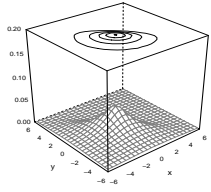}
                \caption{$\boldsymbol\alpha^\prime = (-3,-3)$}
                \label{Dens_SkewDoubleExp_n3.n3}
        \end{subfigure}%
        \begin{subfigure}[b]{0.5\textwidth}
                \includegraphics[height=3.5cm,width=7cm]{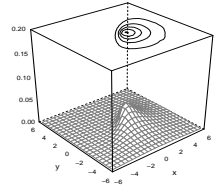}
                \caption{$\boldsymbol\alpha^\prime = (3,-3)$}
                \label{Dens_SkewDoubleExp_3.n3}
        \end{subfigure}
        \caption{Density functions of the bivariate skew double exponential with $\boldsymbol\xi = (0,0)$ and scale matrix $\boldsymbol\Omega = \boldsymbol I_2$, for different shape vectors.}
        \label{Fig:SkewDoubleExp}
\end{figure}

Figure \ref{Fig:SkewDoubleExp} contains the plots of the density functions for the skewed bivariate double exponential variable; we can observe the effect the slant parameter $\boldsymbol\alpha$ has on the shape of the densities as well as the contoured plots obtained after injection of different asymmetries. Simple comparison with the plots of Figure \ref{Fig:SkewT} shows two different ways to account both for skewness and tail weight behaviour.

The moments of the mixing variable are
$\displaystyle E(S^k)=\frac{2^{k/2}\Gamma\left(\frac{p}{2}\right)\Gamma(p+k)}{\Gamma(p)\Gamma\left(\frac{p+k}{2}\right)} : k\geq 1$. Using this expression, we obtain that

$$a=E(S)\left(\frac{4}{\pi}E^2(S)-4(p+2)\right)=
4 E(S)\left(\frac{2}{\pi}\frac{p^2\Gamma^2\left(\frac{p}{2}\right)}{\Gamma^2\left(\frac{p+1}{2}\right)}-(p+2)\right),$$

\noindent which gives negative values when $p<5$ and positive values when $p \geq 5$.

In order to check if the moment condition $\displaystyle\frac{4}{\pi}E^2(S)\geq E(S^2)$ holds in this case, we are going to define the function:

$$g(p)=\frac{1}{p+1}\left[\frac{p\Gamma\left(\frac{p}{2}\right)}{\Gamma\left(\frac{p+1}{2}\right)}\right]^2\hspace{0.25cm}\mbox{:}\hspace{0.25cm}p\geq 1.$$

After taking logarithms, we can see that its first derivative is given by

$$g^\prime (p)=g(p)\left[\frac{2}{p}+\psi\left(\frac{p}{2}\right)-\psi\left(\frac{p+1}{2}\right)-\frac{1}{p+1}\right],$$

\noindent where $\displaystyle\psi(x)=\frac{\Gamma^\prime (x)}{\Gamma(x)}$ is the digamma function.

Taking into account the well-known property: $\displaystyle\psi(x+1)=\frac{1}{x}+\psi(x)$ and the following inequalities regarding the digamma function:
$\displaystyle\log\left(x-\frac{1}{2}\right)<\psi(x)<\log(x)-\frac{1}{2x}$ when $\displaystyle x>\frac{1}{2}$ \citep{Merkle1998}, we get

$$g^\prime (p)=g(p)\left[\psi\left(\frac{p+2}{2}\right)-\psi\left(\frac{p+1}{2}\right)-\frac{1}{p+1}\right]$$

$$>g(p)\left[\log\left(\frac{p+1}{2}\right)-\log\left(\frac{p+1}{2}\right)+\frac{1}{p+1}-\frac{1}{p+1}\right]=0,$$

\noindent which implies that $g(p)$ is a non decreasing function for $p\geq 1$. Consequently, $\displaystyle g(p)\geq g(1)=\frac{\pi}{2}$ or equivalently
$\displaystyle\frac{2}{\pi}\left[\frac{p\Gamma\left(\frac{p}{2}\right)}{\Gamma\left(\frac{p+1}{2}\right)}\right]^2\geq p+1$. This inequality leads to the moment condition $\displaystyle\frac{4}{\pi}E^2(S)\geq E(S^2)$, so the result of Theorem \ref{TheMaxProjectSkew} is also valid for multivariate double exponential vectors.

\subsubsection{The multivariate skew-slash distribution}
\label{Ex.Skew.Slash}

Another flexible model that combines both asymmetry and tail weight behavior is the multivariate skew-slash (SSL) distribution \citep{WangGenton2006}. The multivariate SSL distribution corresponds to the case where the mixing variable is $S=U^{-1/q}$ with $U\sim U(0,1)$ and $q$ a tail weight parameter such that $q>0$. We use the notation $\boldsymbol X \sim SSL_p(\boldsymbol\xi,\boldsymbol\Omega,\boldsymbol\alpha,q)$ to indicate that $\boldsymbol X$ follows a $p$-dimensional SSL distribution with location $\boldsymbol\xi$, scale matrix $\boldsymbol\Omega=\boldsymbol\omega\bar{\boldsymbol\Omega}\boldsymbol\omega$, shape skewness vector $\boldsymbol\alpha$ and tail weight parameter $q>0$. Note that when $q\rightarrow\infty$, the SSL becomes a $p$-dimensional SN variable, i.e. $\boldsymbol X \sim SN_p(\boldsymbol\xi,\boldsymbol\Omega,\boldsymbol\alpha)$.

\begin{figure}[h]
        \centering
        \begin{subfigure}[b]{0.5\textwidth}
                \includegraphics[height=3.5cm,width=7cm]{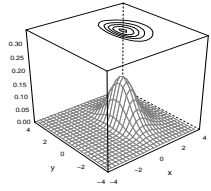}
                \caption{$\boldsymbol\alpha^\prime = (3,0)$}
                \label{Dens_SkewSlash_3.0}
        \end{subfigure}%
        \begin{subfigure}[b]{0.5\textwidth}
                \includegraphics[height=3.5cm,width=7cm]{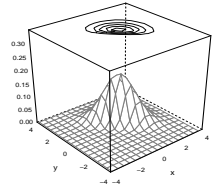}
                \caption{$\boldsymbol\alpha^\prime = (3,3)$}
                \label{Dens_SkewSlash_3.3}
        \end{subfigure}
        \\
        \begin{subfigure}[b]{0.5\textwidth}
                \includegraphics[height=3.5cm,width=7cm]{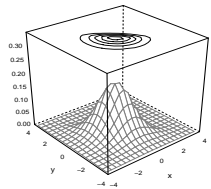}
                \caption{$\boldsymbol\alpha^\prime = (-3,-3)$}
                \label{Dens_SkewSlash_n3.n3}
        \end{subfigure}%
        \begin{subfigure}[b]{0.5\textwidth}
                \includegraphics[height=3.5cm,width=7cm]{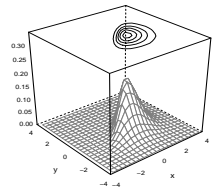}
                \caption{$\boldsymbol\alpha^\prime = (3,-3)$}
                \label{Dens_SkewSlash_3.n3}
        \end{subfigure}
        \caption{Density functions of the bivariate skew-slash variable, with location $\boldsymbol\xi = (0,0)$, scale matrix $\boldsymbol\Omega = \boldsymbol I_2$ and tail weight parameter $q=5$, for different shape vectors.}
        \label{Fig:SkewSlash}
\end{figure}

Figure \ref{Fig:SkewSlash} shows the plots of the bivariate SSL density functions for several directions of asymmetry. The contoured curves are also displayed for each case; they inform us about the different shapes of the scatter plots we would obtain by injecting asymmetry through different directions.

We know that $\displaystyle E(S^m)=E(U^{-m/q})=\frac{q}{q-m}$ for $q>m$; here we assume that $q>3$. Let us define the function:

$$g(q)=\frac{E(S^2)}{E(S)^2}=\frac{(q-1)^2}{q(q-2)}\hspace{0.25cm}\mbox{:}\hspace{0.25cm}q>3.$$

Since $g$ is a decreasing function we have $\displaystyle g(q)\leq g(4)\leq\frac{4}{\pi}$; consequently, the moment condition assumed by Theorem \ref{TheMaxProjectSkew} holds.

\section{Simulation experiment}
\label{Numerical_work}

In this section we shed light on the previous theoretical findings by means of a simulation experiment for artificial data. The computation of the maximal skewness direction $\boldsymbol\eta$ is carried out using the functions implemented in the \textsf{MaxSkew} R package \citep{MaxSkew}.

Now, we confine ourselves to the multivariate skew-t subfamily within the SMSN class of distributions; the multivariate skew-t model is a popular distribution to handle heavy tails and asymmetry deviations from normality in multivariate settings \citep{AzzaliniCapitanio2003}. A simulation experiment for scenarios ranging from distributions with heavy tails to those having nearly normal tails, as regulated by the tail-weight parameter $\nu$, is carried out by drawing $5000$ samples from a skew-t distribution with the following parameter settings: $\boldsymbol\xi =\boldsymbol 0$, $\boldsymbol{\bar{\Omega}}$ a correlation Toeplitz matrix defined by  $\boldsymbol{\bar{\Omega}}=(\omega_{i,j})_{1\leq i,j\leq p}$, where $\omega_{i,j}=\rho^{|i-j|} : 1\leq i\leq j\leq p$ for $\rho= -0.8, -0.3, 0.4, 0.9$ and $\boldsymbol \omega$ a diagonal matrix whose entries are generated at random from the integer values between $1$ and $5$. The simulation experiment is repeated for different dimensions of the input vector $p=2, 10, 18$, tail-weight parameters $\nu=4, 8, 20, 100$ and sample sizes $n=20, 100, 500$. For each case, the maximal skewness direction was calculated using the {\em Singular Value Decomposition} of the third order moment matrix of the model, as implemented by the \textsf{MaxSkew} R package \citep{MaxSkew}, and the skewness coefficient of the projected data is computed accordingly. The mean square errors (MSE) are calculated by comparison with the exact value given by the theory.

\begin{table}
\centering
\begin{tabular}{||c|c|cccc||}
\hline\hline
$p$ & ${\Large n\diagdown \nu }$ & \textit{4} & \textit{8} &
\textit{20} & \textit{100} \\ \hline
& 20 & \multicolumn{1}{|r}{18.3301} & \multicolumn{1}{r}{2.5141} &
\multicolumn{1}{r}{1.1359} & \multicolumn{1}{r||}{0.8819} \\
\textit{2} & 100 & \multicolumn{1}{|r}{49.6732} & \multicolumn{1}{r}{3.0758}
& \multicolumn{1}{r}{0.2508} & \multicolumn{1}{r||}{0.0731} \\
& 500 & \multicolumn{1}{|r}{305.0978} & \multicolumn{1}{r}{0.7389} &
\multicolumn{1}{r}{0.0237} & \multicolumn{1}{r||}{0.0041} \\ \hline
& 20 & \multicolumn{1}{|r}{22.995} & \multicolumn{1}{r}{48.4658} &
\multicolumn{1}{r}{40.6828} & \multicolumn{1}{r||}{36.606} \\
\textit{10} & 100 & \multicolumn{1}{|r}{405.0933} & \multicolumn{1}{r}{61.763
} & \multicolumn{1}{r}{5.8845} & \multicolumn{1}{r||}{1.3423} \\
& 500 & \multicolumn{1}{|r}{2591.437} & \multicolumn{1}{r}{40.7776} &
\multicolumn{1}{r}{0.2902} & \multicolumn{1}{r||}{0.049} \\ \hline
& 20 & \multicolumn{1}{|r}{44.8858} & \multicolumn{1}{r}{39.79} &
\multicolumn{1}{r}{46.4198} & \multicolumn{1}{r||}{49.2469} \\
\textit{18} & 100 & \multicolumn{1}{|r}{959.665} & \multicolumn{1}{r}{
277.2378} & \multicolumn{1}{r}{37.0411} & \multicolumn{1}{r||}{8.1573} \\
& 500 & \multicolumn{1}{|r}{7934.145} & \multicolumn{1}{r}{138.4876} &
\multicolumn{1}{r}{1.4523} & \multicolumn{1}{r||}{0.09} \\ \hline\hline
\end{tabular}
\caption{MSE obtained from the simulations when $\rho =-0.80$.}
\label{TableRhoM080}
\end{table}

\begin{table}
\centering
\begin{tabular}{||c|c|cccc||}
\hline\hline
$p$ & ${\Large n\diagdown \nu }$ & \textit{4} & \textit{8} &
\textit{20} & \textit{100} \\ \hline
& 20 & \multicolumn{1}{|r}{9.8158} & \multicolumn{1}{r}{2.5211} &
\multicolumn{1}{r}{1.2035} & \multicolumn{1}{r||}{0.7772} \\
\textit{2} & 100 & \multicolumn{1}{|r}{47.4628} & \multicolumn{1}{r}{2.1425}
& \multicolumn{1}{r}{0.2} & \multicolumn{1}{r||}{0.0636} \\
& 500 & \multicolumn{1}{|r}{221.6212} & \multicolumn{1}{r}{0.7048} &
\multicolumn{1}{r}{0.0163} & \multicolumn{1}{r||}{0.0043} \\ \hline
& 20 & \multicolumn{1}{|r}{34.1029} & \multicolumn{1}{r}{55.987} &
\multicolumn{1}{r}{44.5085} & \multicolumn{1}{r||}{38.712} \\
\textit{10} & 100 & \multicolumn{1}{|r}{525.8873} & \multicolumn{1}{r}{
68.5109} & \multicolumn{1}{r}{6.5594} & \multicolumn{1}{r||}{1.5701} \\
& 500 & \multicolumn{1}{|r}{2918.442} & \multicolumn{1}{r}{22.2143} &
\multicolumn{1}{r}{0.2687} & \multicolumn{1}{r||}{0.0263} \\ \hline
& 20 & \multicolumn{1}{|r}{20.2875} & \multicolumn{1}{r}{45.2771} &
\multicolumn{1}{r}{51.4088} & \multicolumn{1}{r||}{53.5157} \\
\textit{18} & 100 & \multicolumn{1}{|r}{1235.183} & \multicolumn{1}{r}{
271.2368} & \multicolumn{1}{r}{40.7157} & \multicolumn{1}{r||}{8.7655} \\
& 500 & \multicolumn{1}{|r}{8450.464} & \multicolumn{1}{r}{175.6163} &
\multicolumn{1}{r}{1.5504} & \multicolumn{1}{r||}{0.0692} \\ \hline\hline
\end{tabular}
\caption{MSE obtained from the simulations when $\rho =-0.30$.}
\label{TableRhoM030}
\end{table}

\begin{table}
\centering
\begin{tabular}{||c|c|cccc||}
\hline\hline
$p$ & ${\Large n\diagdown \nu }$ & \textit{4} & \textit{8} &
\textit{20} & \textit{100} \\ \hline
& 20 & \multicolumn{1}{|r}{12.907} & \multicolumn{1}{r}{2.6373} &
\multicolumn{1}{r}{1.2045} & \multicolumn{1}{r||}{0.7696} \\
\textit{2} & 100 & \multicolumn{1}{|r}{60.3853} & \multicolumn{1}{r}{2.6143}
& \multicolumn{1}{r}{0.2034} & \multicolumn{1}{r||}{0.0612} \\
& 500 & \multicolumn{1}{|r}{387.187} & \multicolumn{1}{r}{0.8915} &
\multicolumn{1}{r}{0.0167} & \multicolumn{1}{r||}{0.004} \\ \hline
& 20 & \multicolumn{1}{|r}{28.7548} & \multicolumn{1}{r}{55.5959} &
\multicolumn{1}{r}{43.6206} & \multicolumn{1}{r||}{38.5087} \\
\textit{10} & 100 & \multicolumn{1}{|r}{463.0498} & \multicolumn{1}{r}{
64.6375} & \multicolumn{1}{r}{6.5316} & \multicolumn{1}{r||}{1.5479} \\
& 500 & \multicolumn{1}{|r}{3315.722} & \multicolumn{1}{r}{30.5471} &
\multicolumn{1}{r}{0.2511} & \multicolumn{1}{r||}{0.0266} \\ \hline
& 20 & \multicolumn{1}{|r}{19.9435} & \multicolumn{1}{r}{46.5429} &
\multicolumn{1}{r}{50.523} & \multicolumn{1}{r||}{51.6388} \\
\textit{18} & 100 & \multicolumn{1}{|r}{1183.423} & \multicolumn{1}{r}{
286.0585} & \multicolumn{1}{r}{39.3395} & \multicolumn{1}{r||}{8.3242} \\
& 500 & \multicolumn{1}{|r}{8341.767} & \multicolumn{1}{r}{174.696} &
\multicolumn{1}{r}{1.6612} & \multicolumn{1}{r||}{0.0613} \\ \hline\hline
\end{tabular}
\caption{MSE obtained from the simulations when $\rho =0.40$.}
\label{TableRho040}
\end{table}

\begin{table}
\centering
\begin{tabular}{||c|c|cccc||}
\hline\hline
$p$ & ${\Large n\diagdown \nu }$ & \textit{4} & \textit{8} &
\textit{20} & \textit{100} \\ \hline
& 20 & \multicolumn{1}{|r}{19.2307} & \multicolumn{1}{r}{2.6322} &
\multicolumn{1}{r}{1.2748} & \multicolumn{1}{r||}{0.7573} \\
\textit{2} & 100 & \multicolumn{1}{|r}{61.9649} & \multicolumn{1}{r}{2.1586}
& \multicolumn{1}{r}{0.1788} & \multicolumn{1}{r||}{0.0623} \\
& 500 & \multicolumn{1}{|r}{342.9048} & \multicolumn{1}{r}{0.7787} &
\multicolumn{1}{r}{0.0174} & \multicolumn{1}{r||}{0.0056} \\ \hline
& 20 & \multicolumn{1}{|r}{26.628} & \multicolumn{1}{r}{45.0348} &
\multicolumn{1}{r}{40.14} & \multicolumn{1}{r||}{35.4} \\
\textit{10} & 100 & \multicolumn{1}{|r}{363.5182} & \multicolumn{1}{r}{
57.9201} & \multicolumn{1}{r}{6.0117} & \multicolumn{1}{r||}{1.3955} \\
& 500 & \multicolumn{1}{|r}{3019.186} & \multicolumn{1}{r}{24.9832} &
\multicolumn{1}{r}{0.2786} & \multicolumn{1}{r||}{0.0623} \\ \hline
& 20 & \multicolumn{1}{|r}{63.4021} & \multicolumn{1}{r}{36.7311} &
\multicolumn{1}{r}{45.0271} & \multicolumn{1}{r||}{46.7664} \\
\textit{18} & 100 & \multicolumn{1}{|r}{903.8381} & \multicolumn{1}{r}{
257.4515} & \multicolumn{1}{r}{36.6135} & \multicolumn{1}{r||}{7.3076} \\
& 500 & \multicolumn{1}{|r}{7840.981} & \multicolumn{1}{r}{156.9846} &
\multicolumn{1}{r}{1.3074} & \multicolumn{1}{r||}{0.1164} \\ \hline\hline
\end{tabular}
\caption{MSE obtained from the simulations when $\rho =0.90$.}
\label{TableRho090}
\end{table}

\begin{figure}[h]
        \begin{subfigure}[b]{0.5\textwidth}
                \centering
                \includegraphics[height=4cm,width=4cm]{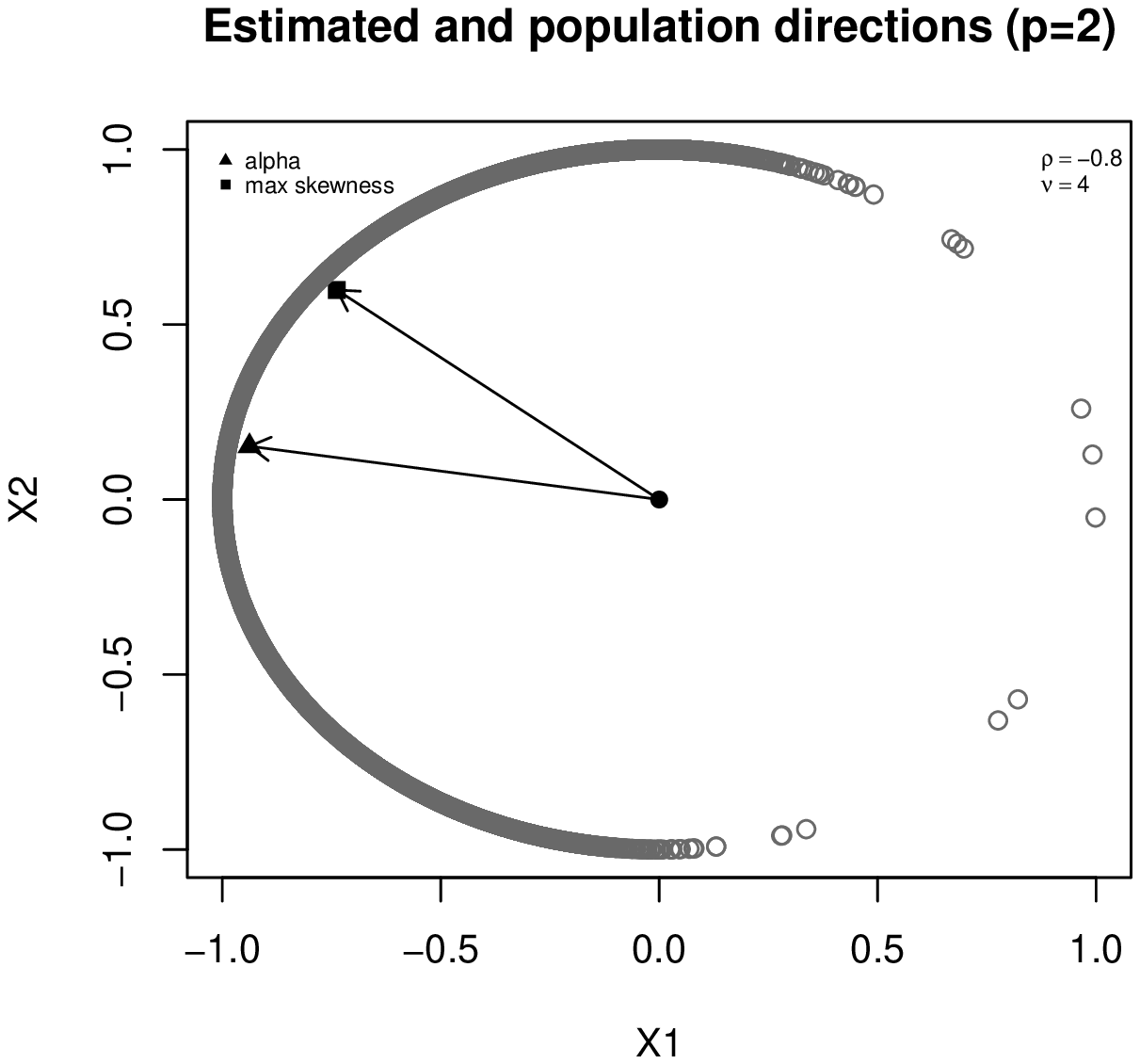}
        \end{subfigure}%
        \begin{subfigure}[b]{0.5\textwidth}
                \centering
                \includegraphics[height=4cm,width=4cm]{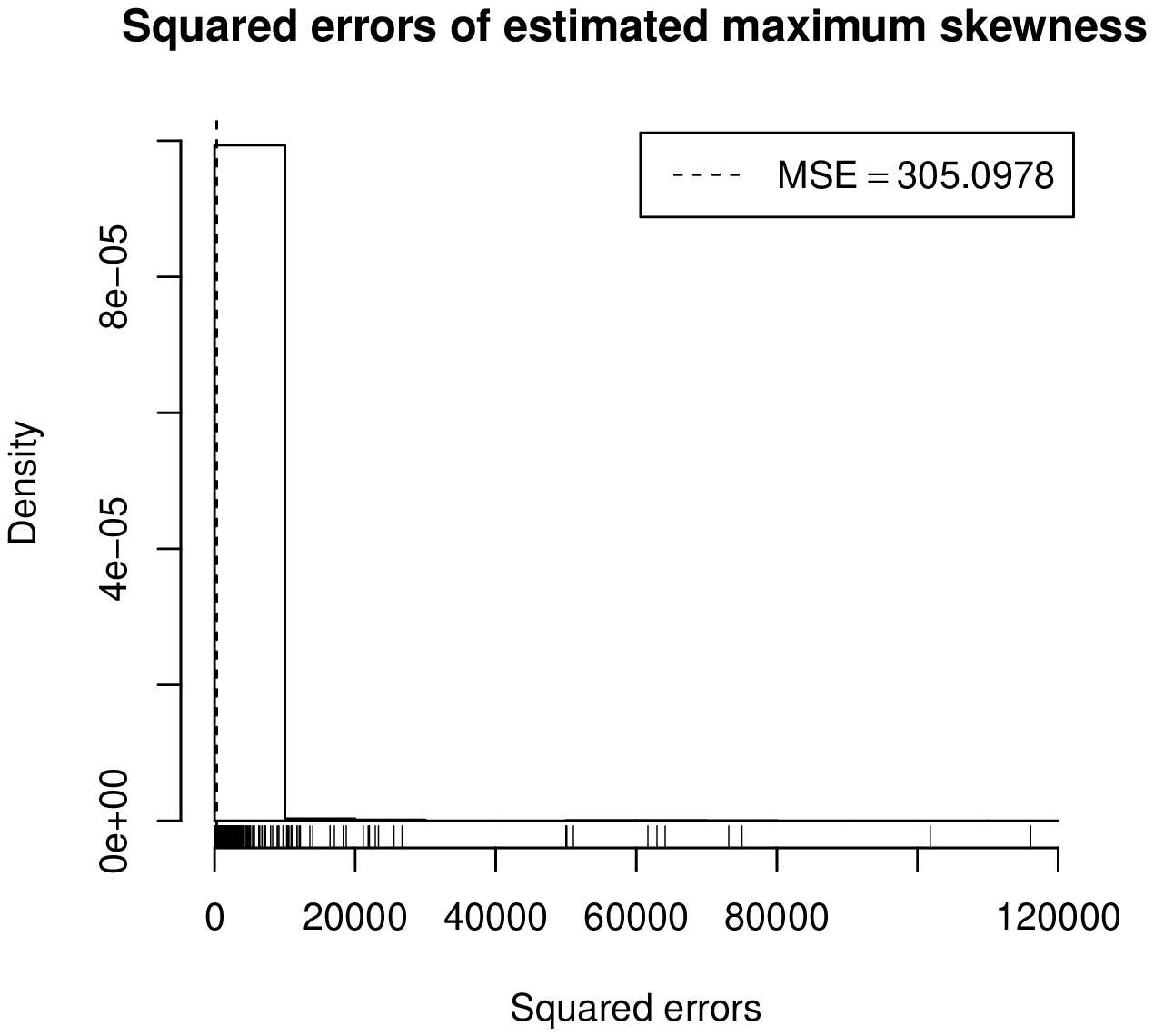}
        \end{subfigure}
        \\
        \begin{subfigure}[b]{0.5\textwidth}
                \centering
                \includegraphics[height=4cm,width=4cm]{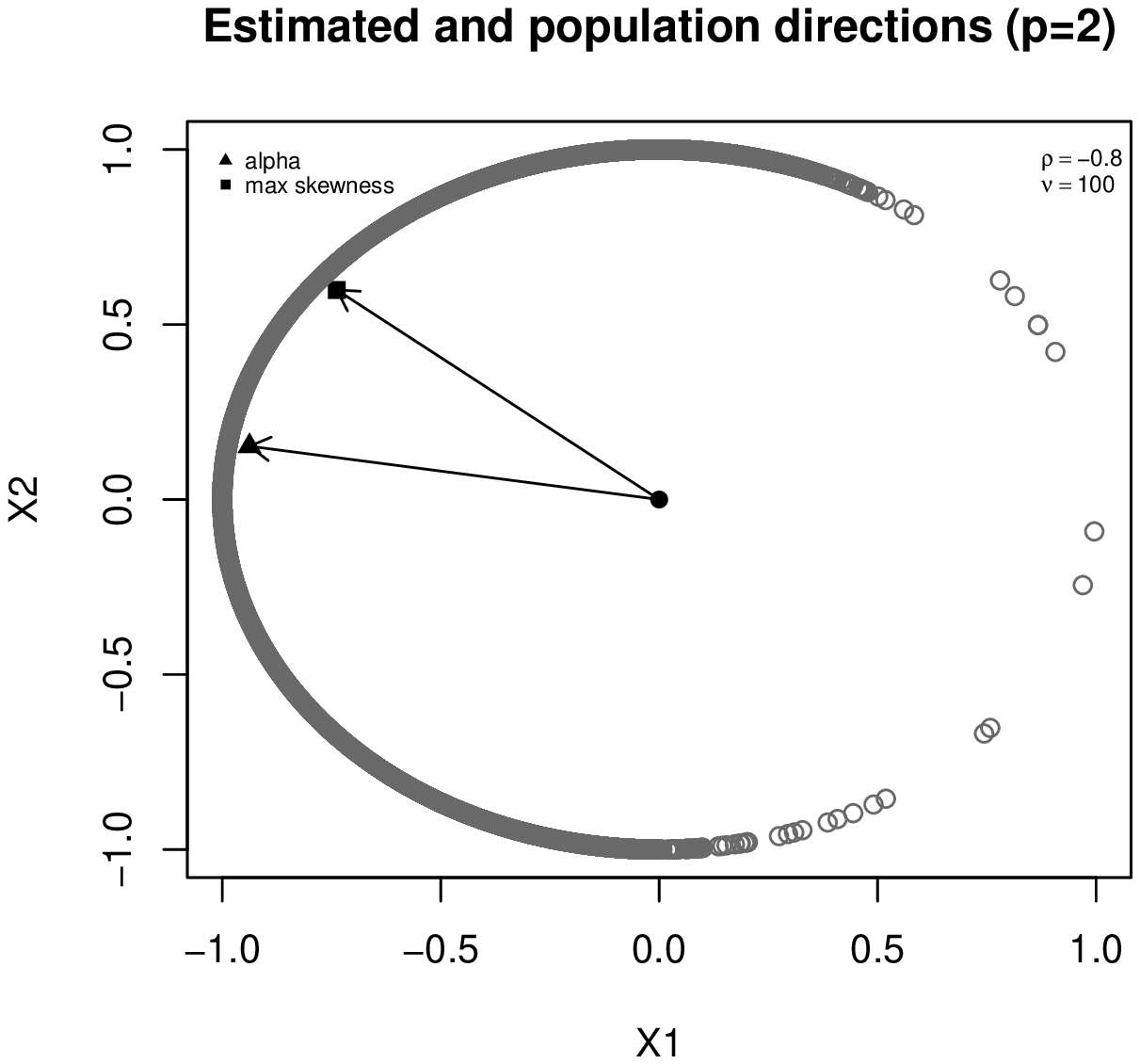}
        \end{subfigure}%
        \begin{subfigure}[b]{0.5\textwidth}
                \centering
                \includegraphics[height=4cm,width=4cm]{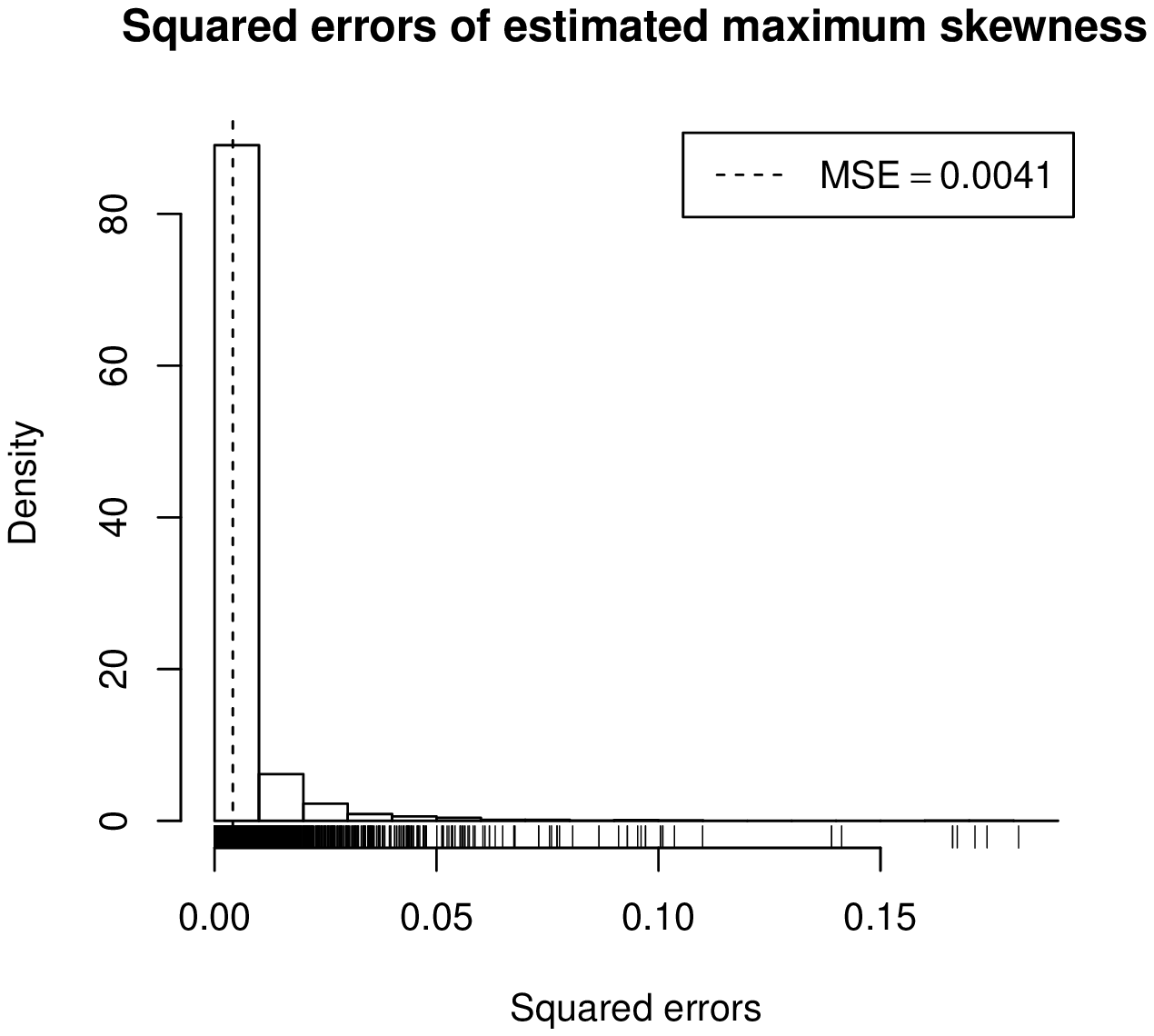}
        \end{subfigure}
        \begin{subfigure}[b]{0.5\textwidth}
                \centering
                \includegraphics[height=4cm,width=4cm]{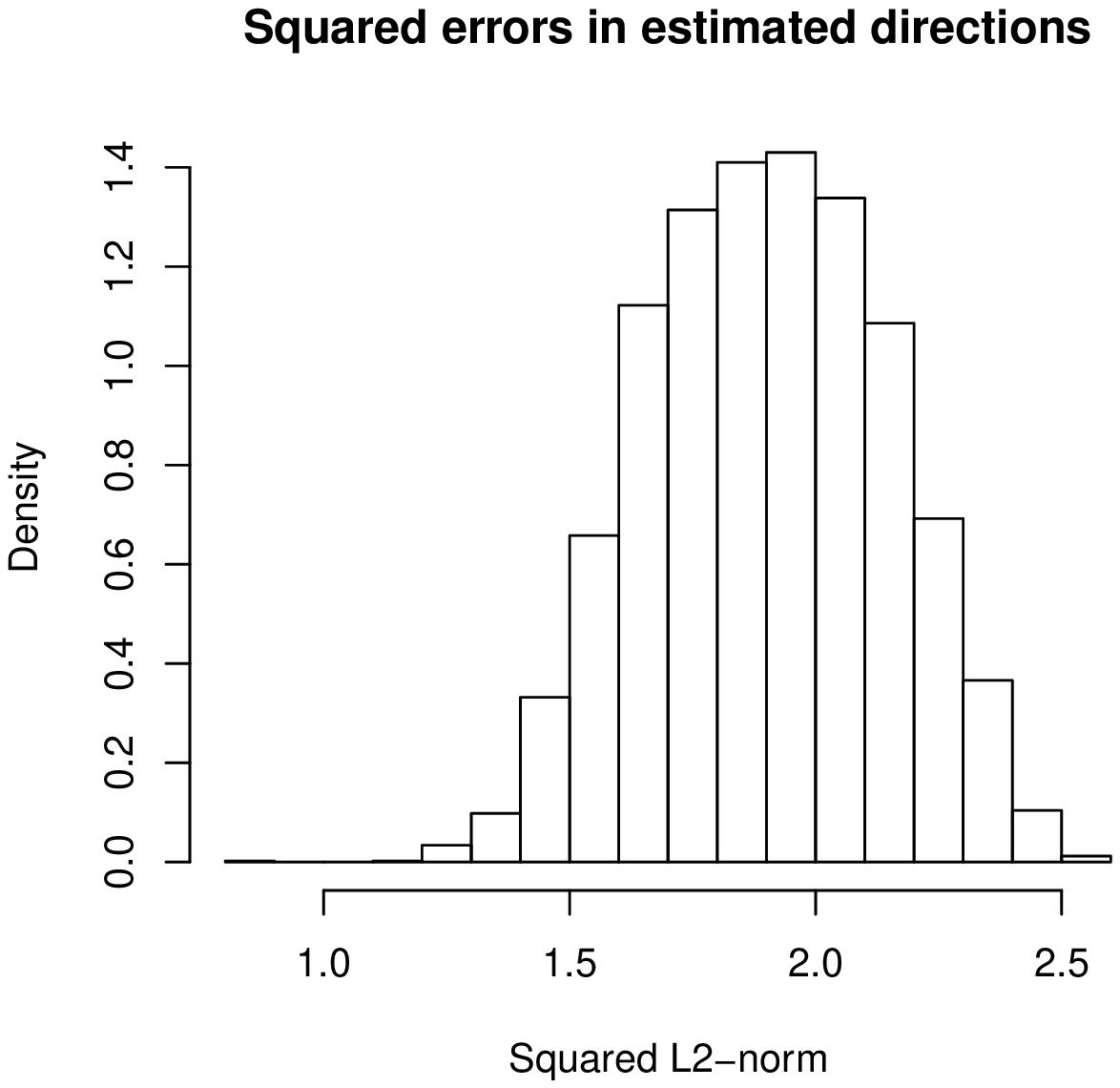}
        \end{subfigure}%
        \begin{subfigure}[b]{0.5\textwidth}
                \centering
                \includegraphics[height=4cm,width=4cm]{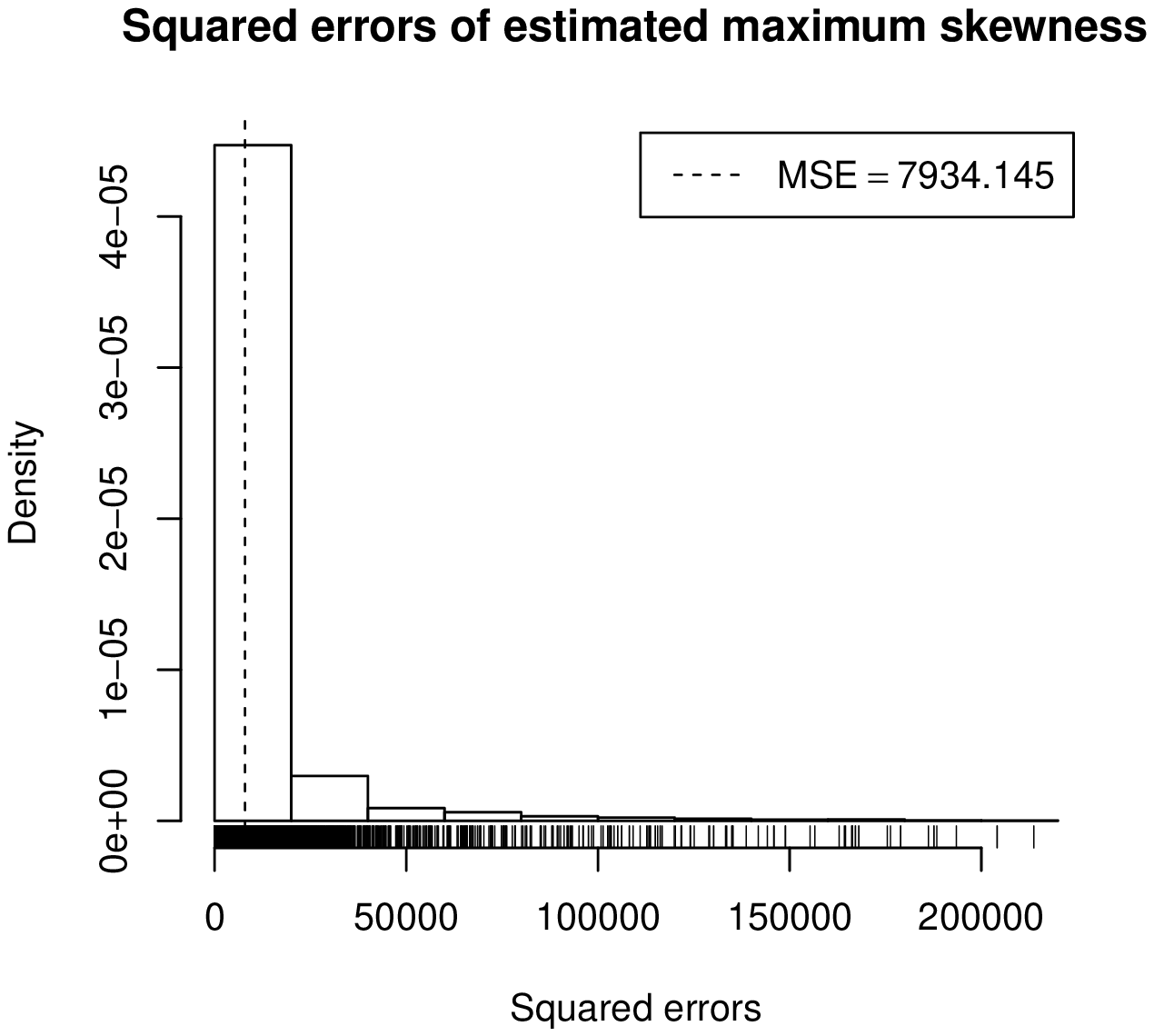}
        \end{subfigure}
        \\
        \begin{subfigure}[b]{0.5\textwidth}
                \centering
                \includegraphics[height=4cm,width=4cm]{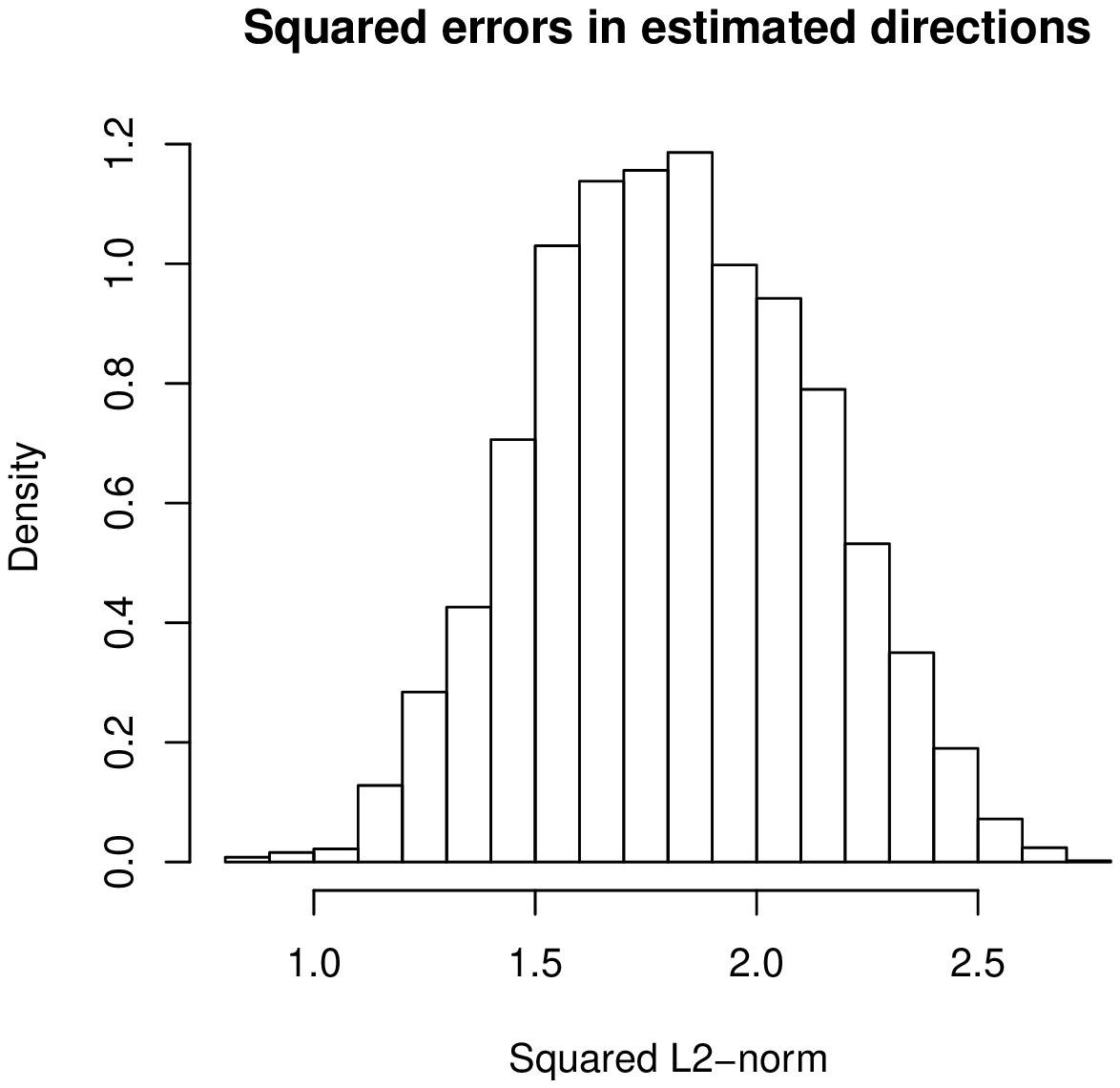}
        \end{subfigure}%
        \begin{subfigure}[b]{0.5\textwidth}
                \centering
                \includegraphics[height=4cm,width=4cm]{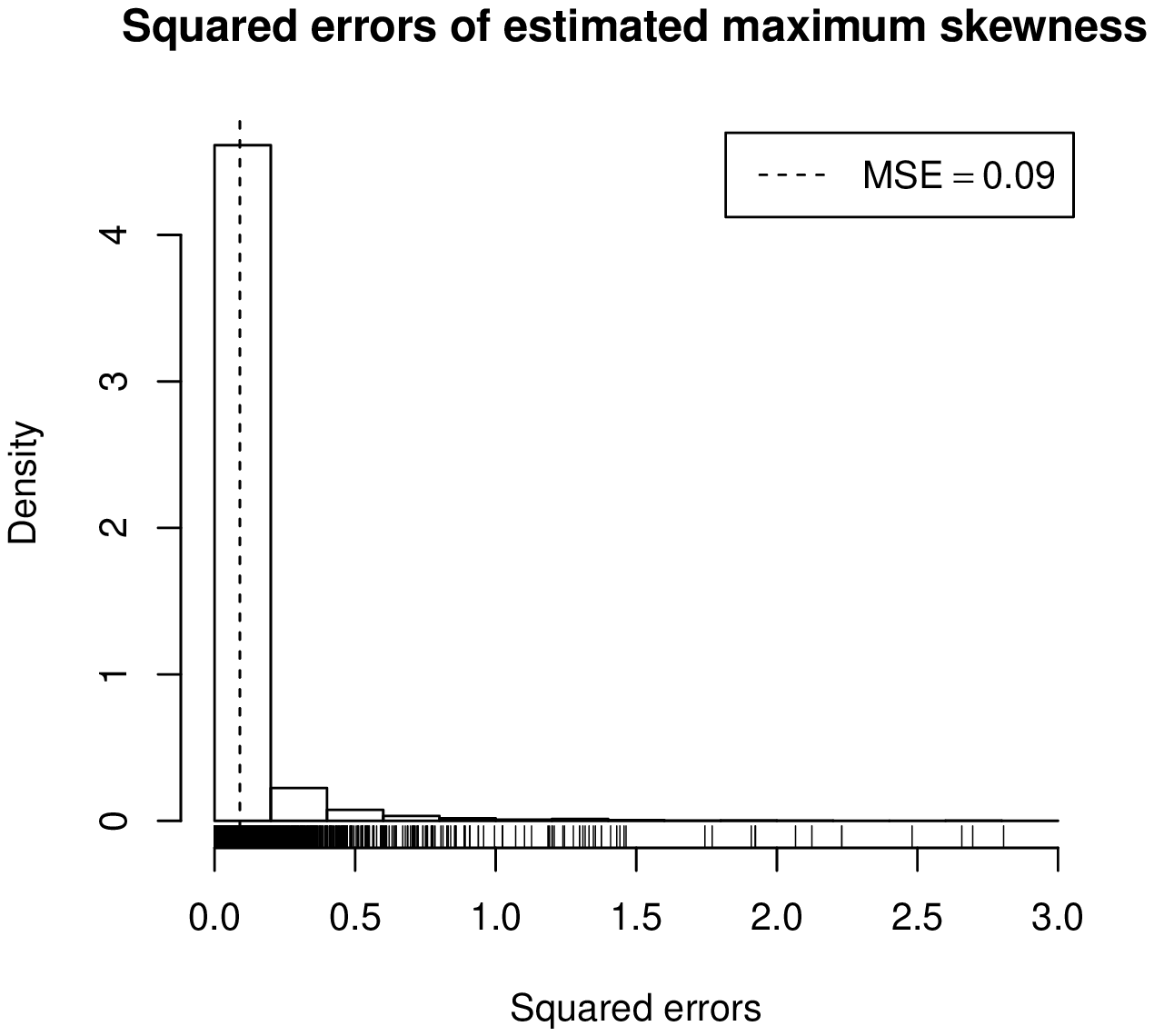}
        \end{subfigure}
        \caption{Plots when $n=500$  in Table \ref{TableRhoM080}: clock plots of the estimated directions and histograms of the squared error in the estimation of the maximal skewness when $p=2, \nu =4$ (first row) and $p=2, \nu =100$ (second row), as well as histograms of the squared $L2$-norm error of the estimated direction and the squared error of the maximal skewness when $p=18, \nu =4$ (third row) and $p=18, \nu =100$ (fourth row).}
        \label{PlotsRhoM080}
\end{figure}

\begin{figure}[h]
         \begin{subfigure}[b]{0.5\textwidth}
                \centering
                \includegraphics[height=4cm,width=4cm]{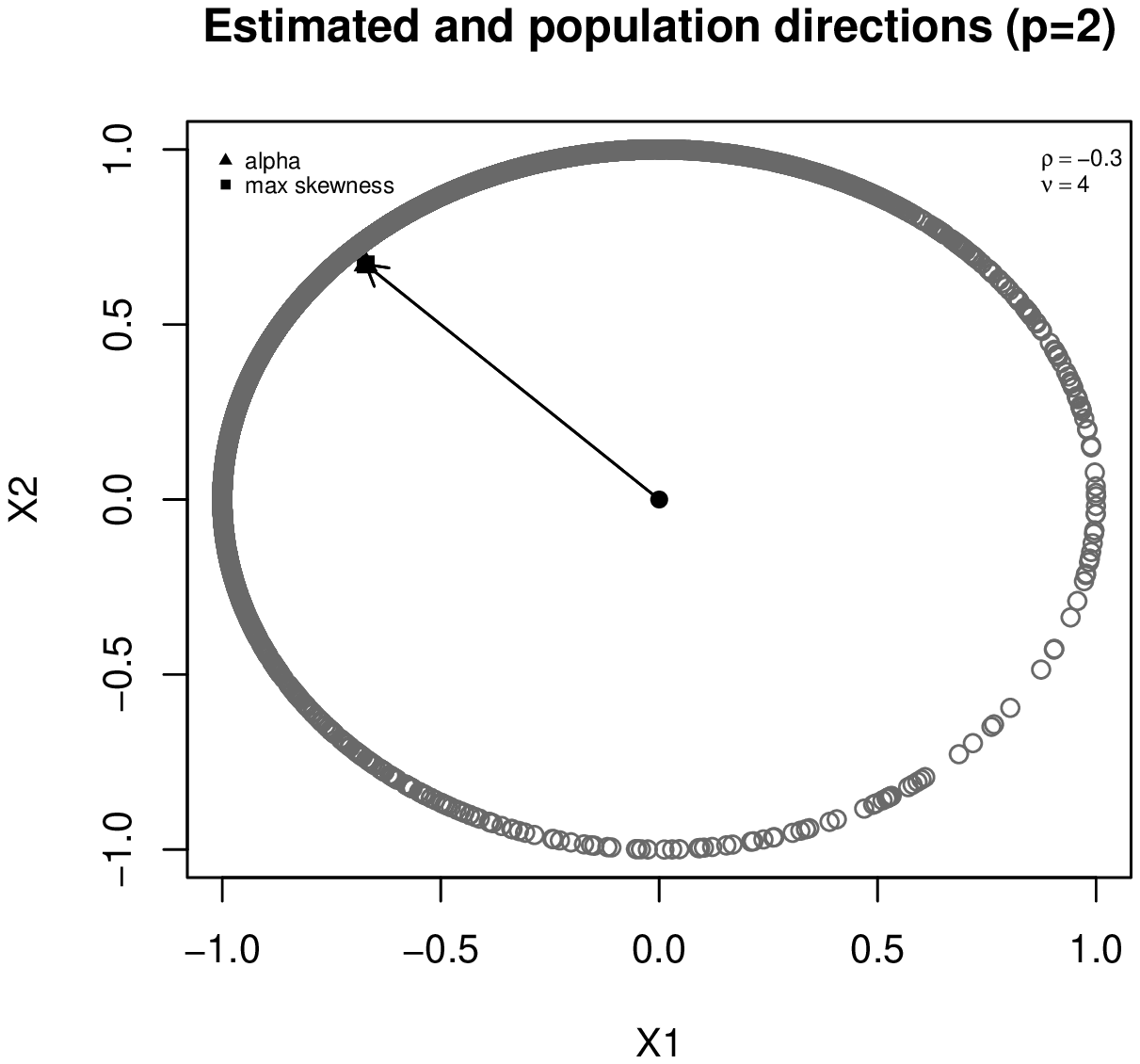}
        \end{subfigure}%
        \begin{subfigure}[b]{0.5\textwidth}
                \centering
                \includegraphics[height=4cm,width=4cm]{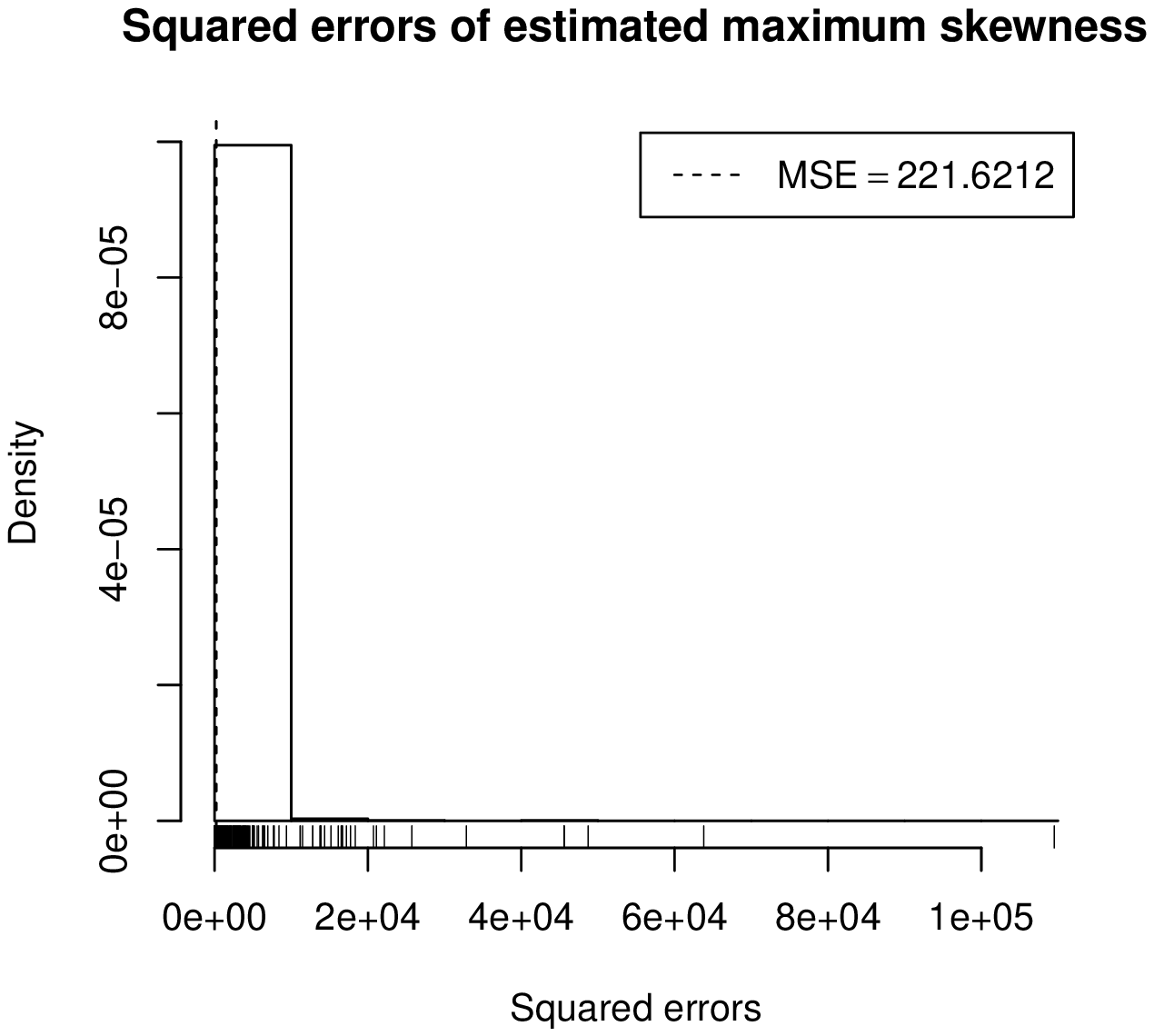}
        \end{subfigure}
        \\
        \begin{subfigure}[b]{0.5\textwidth}
                \centering
                \includegraphics[height=4cm,width=4cm]{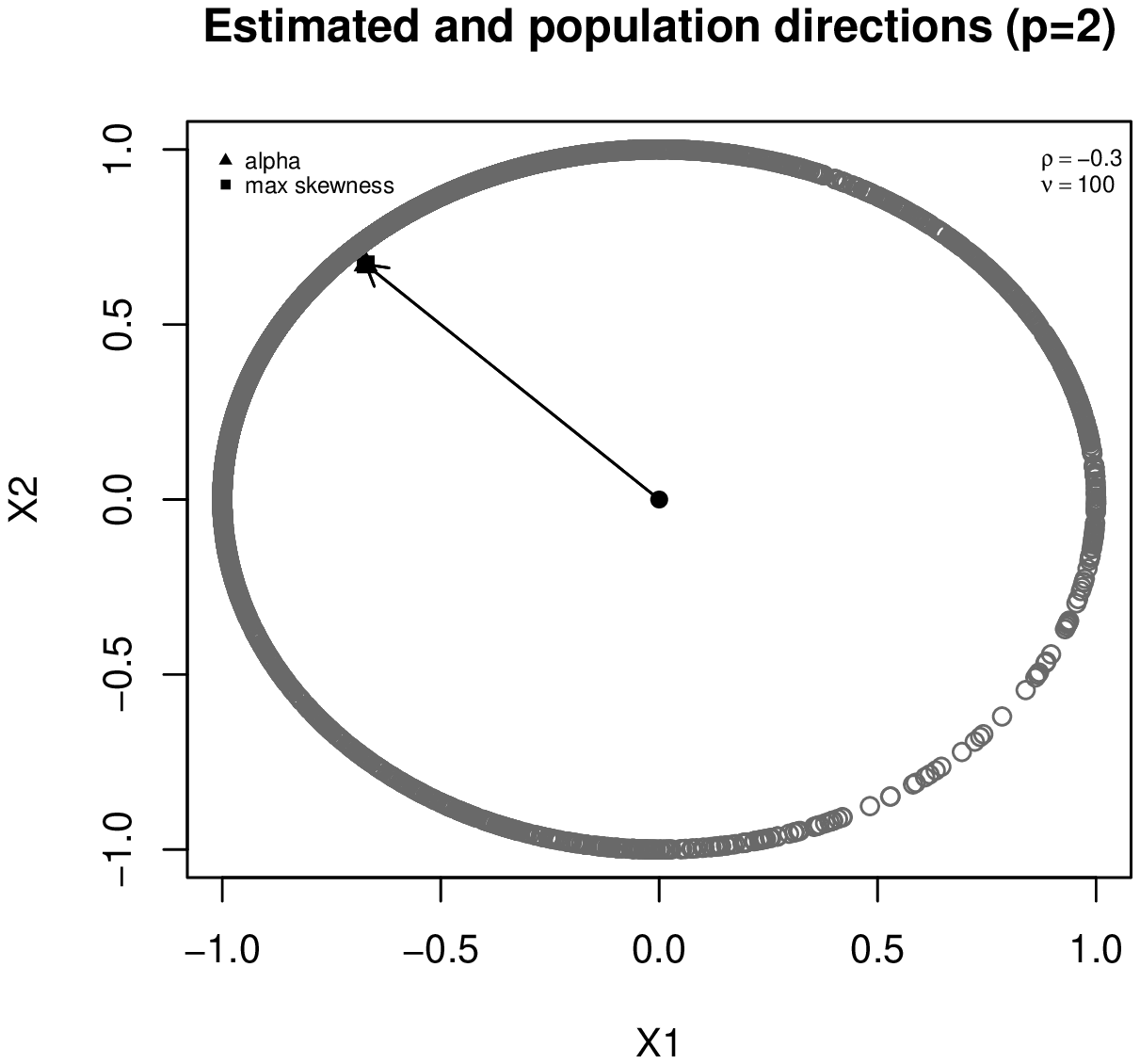}
        \end{subfigure}%
        \begin{subfigure}[b]{0.5\textwidth}
                \centering
                \includegraphics[height=4cm,width=4cm]{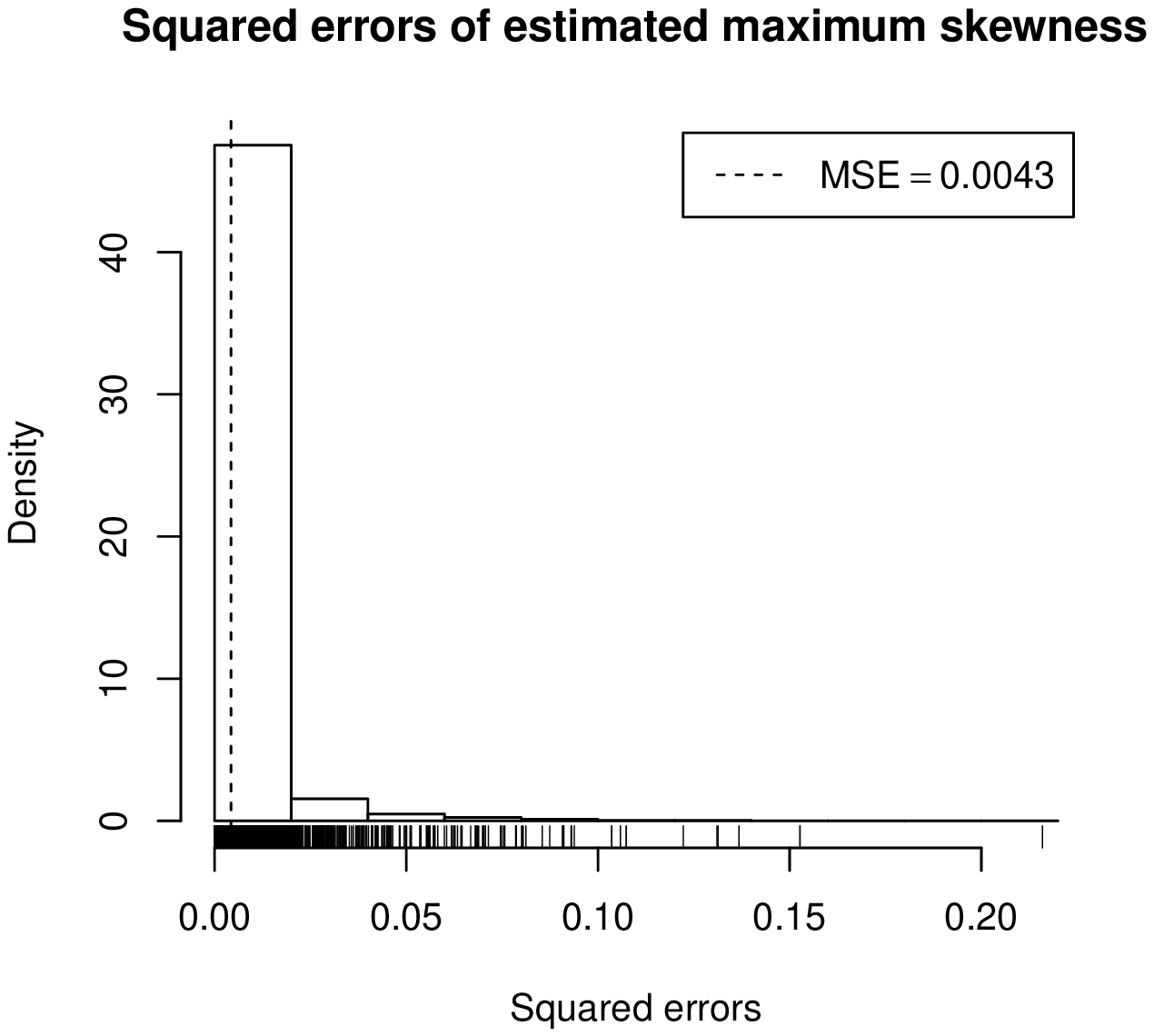}
        \end{subfigure}
        \begin{subfigure}[b]{0.5\textwidth}
                \centering
                \includegraphics[height=4cm,width=4cm]{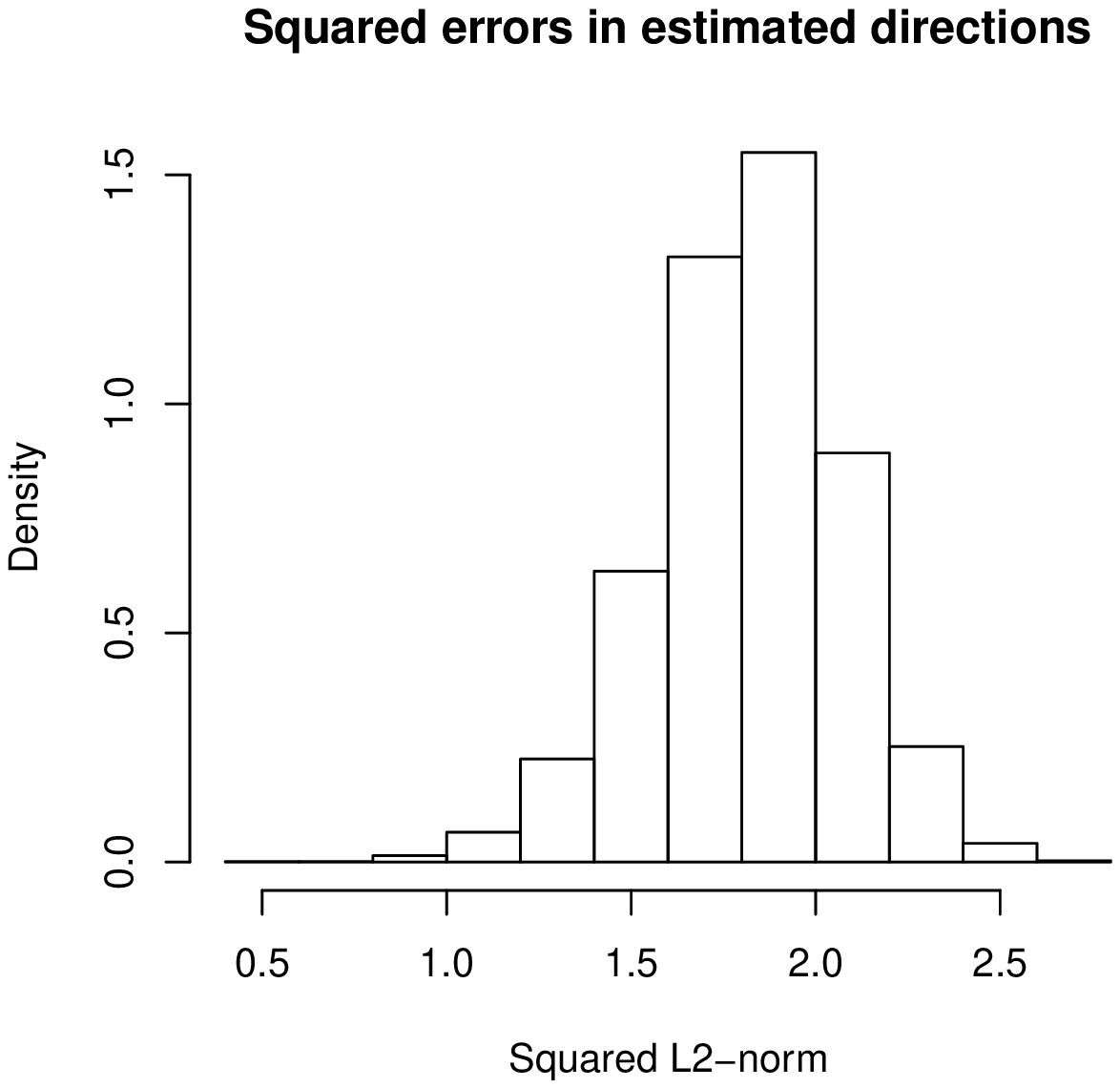}
        \end{subfigure}%
        \begin{subfigure}[b]{0.5\textwidth}
                \centering
                \includegraphics[height=4cm,width=4cm]{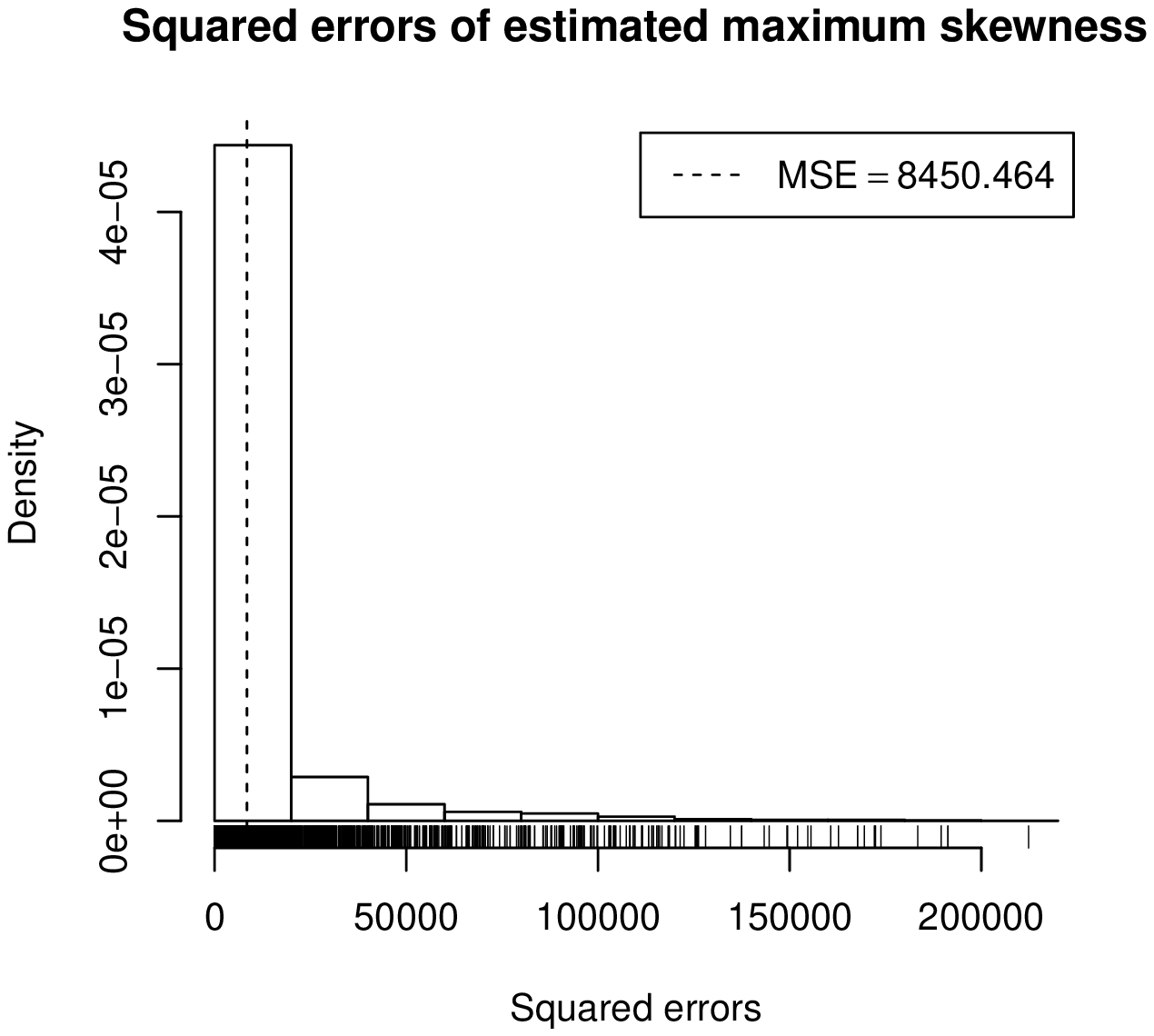}
        \end{subfigure}
        \\
        \begin{subfigure}[b]{0.5\textwidth}
                \centering
                \includegraphics[height=4cm,width=4cm]{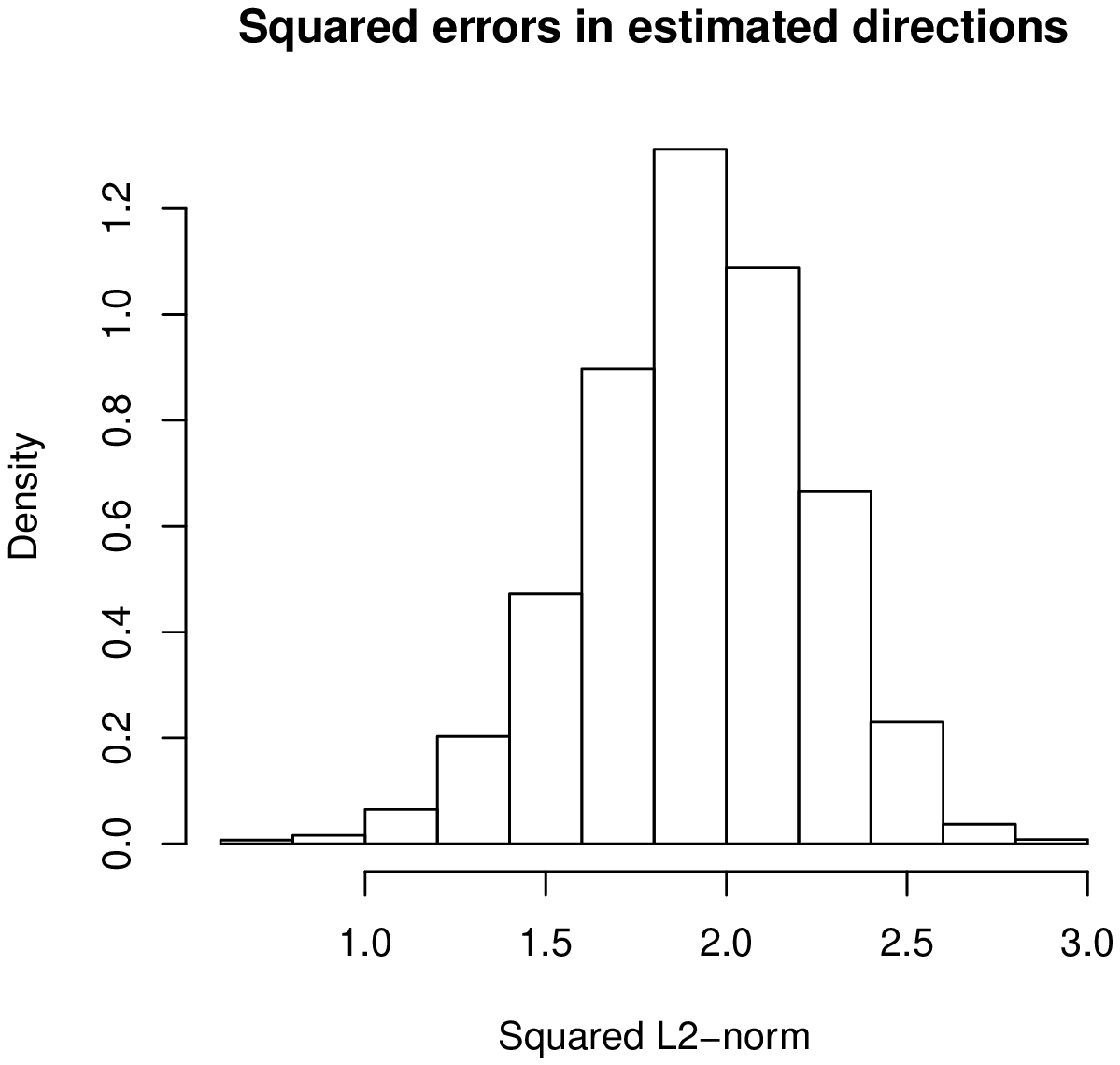}
        \end{subfigure}%
        \begin{subfigure}[b]{0.5\textwidth}
                \centering
                \includegraphics[height=4cm,width=4cm]{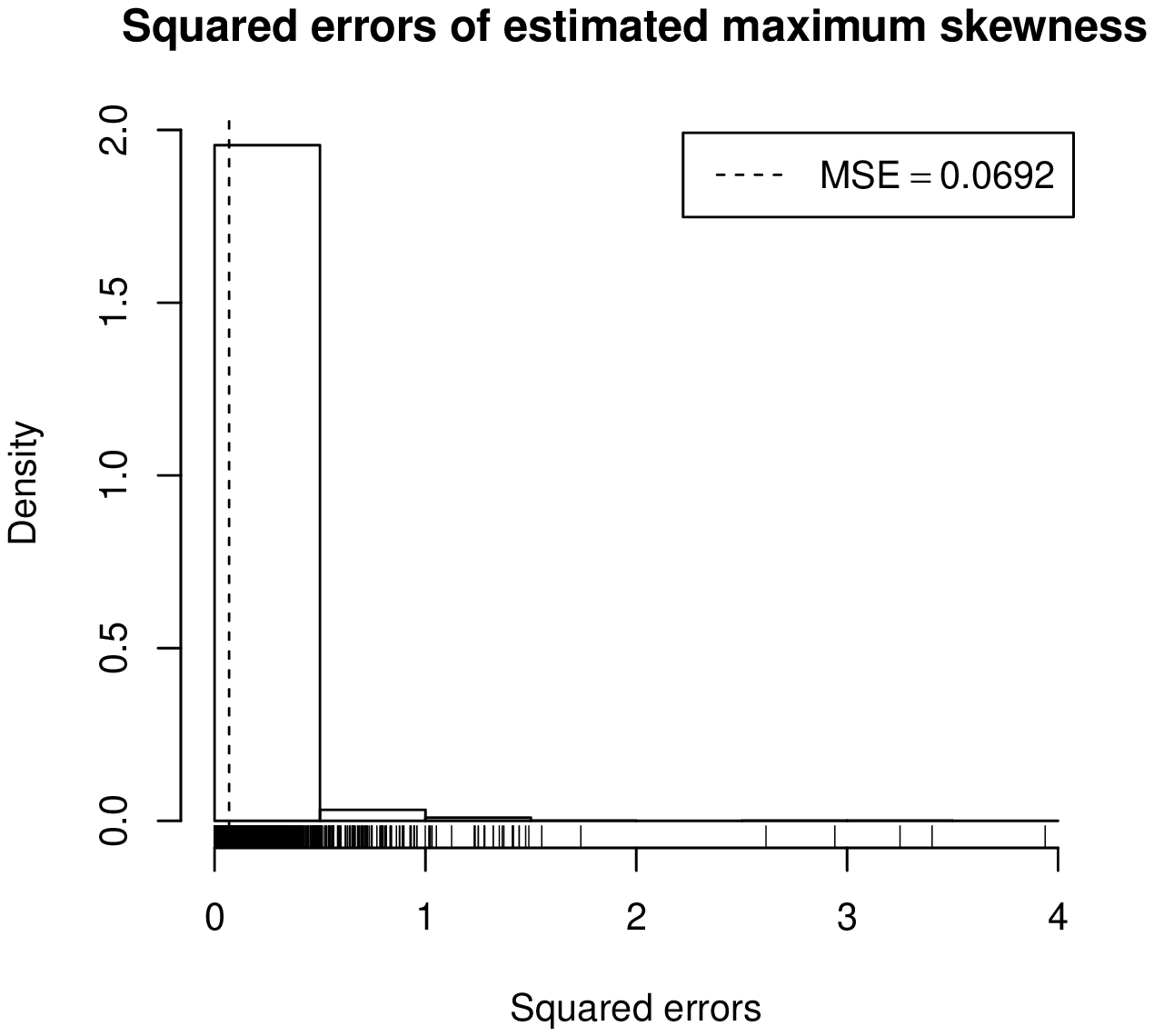}
        \end{subfigure}
        \caption{Plots when $n=500$  in Table \ref{TableRhoM030}: clock plots of the estimated directions and histograms of the squared error in the estimation of the maximal skewness when $p=2, \nu =4$ (first row) and $p=2, \nu =100$ (second row), as well as histograms of the squared $L2$-norm error of the estimated direction and the squared error of the maximal skewness when $p=18, \nu =4$ (third row) and $p=18, \nu =100$ (fourth row).}
        \label{PlotsRhoM030}
\end{figure}

\begin{figure}[h]
         \begin{subfigure}[b]{0.5\textwidth}
                \centering
                \includegraphics[height=4cm,width=4cm]{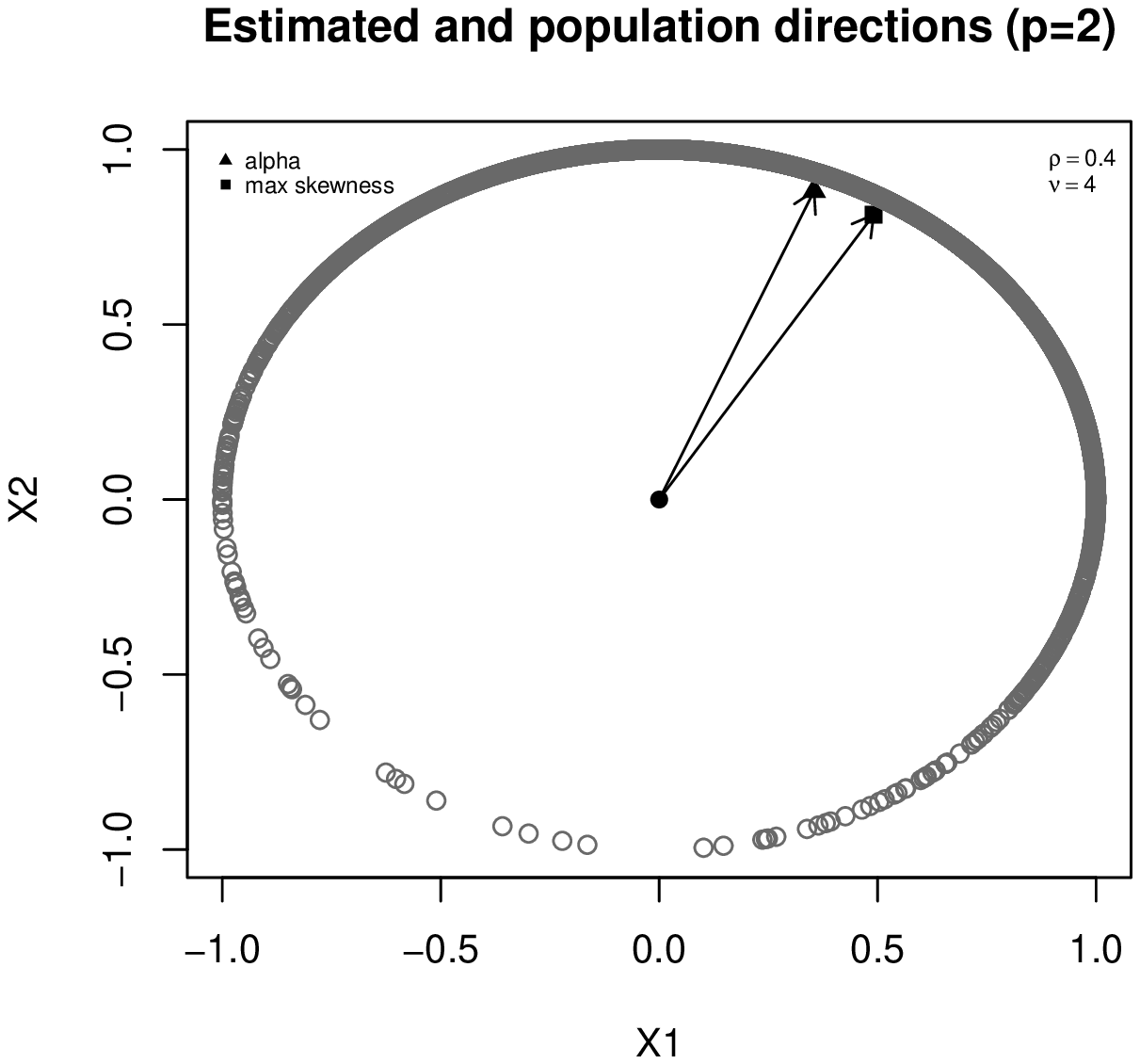}
        \end{subfigure}%
        \begin{subfigure}[b]{0.5\textwidth}
                \centering
                \includegraphics[height=4cm,width=4cm]{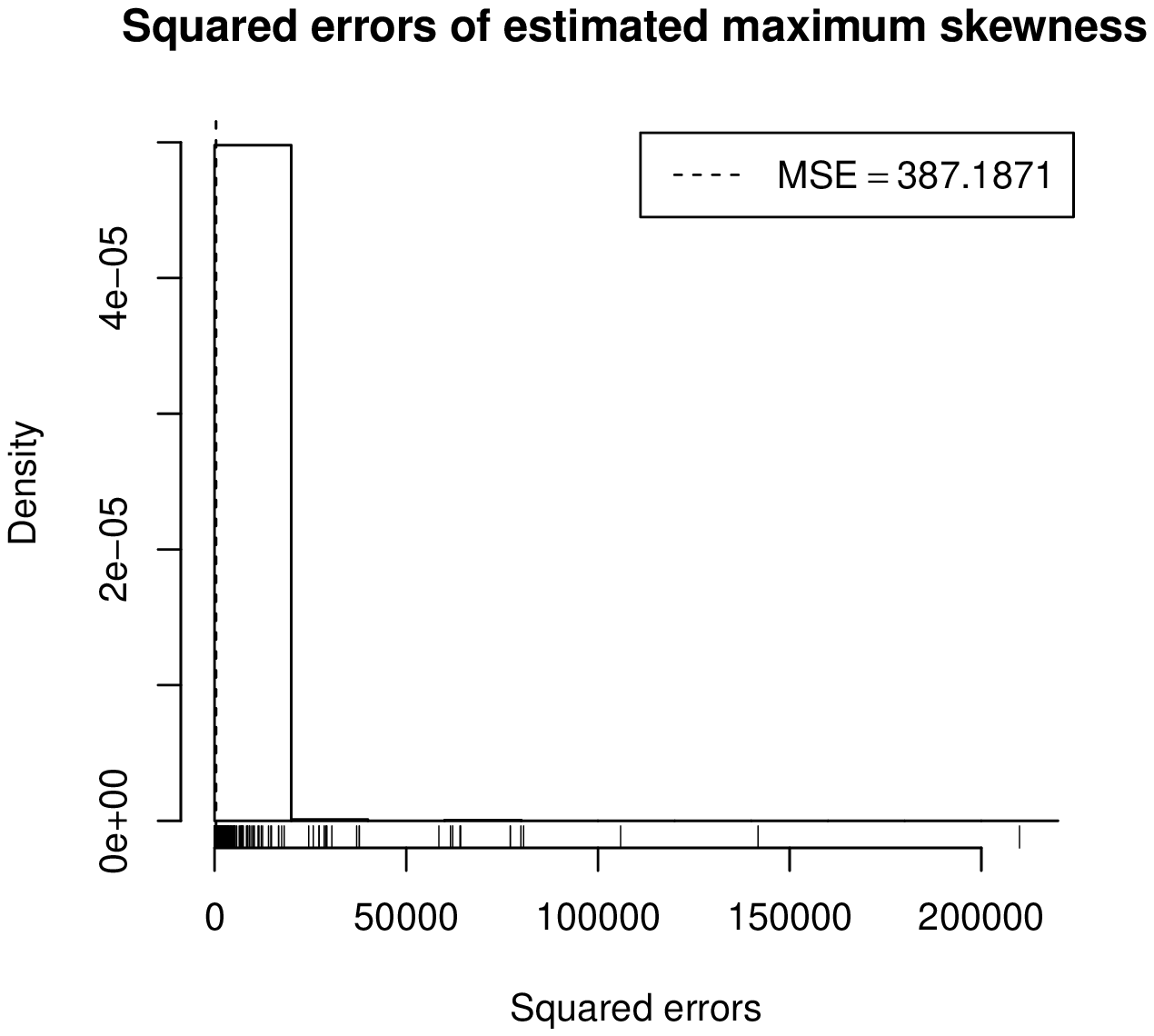}
        \end{subfigure}
        \\
        \begin{subfigure}[b]{0.5\textwidth}
                \centering
                \includegraphics[height=4cm,width=4cm]{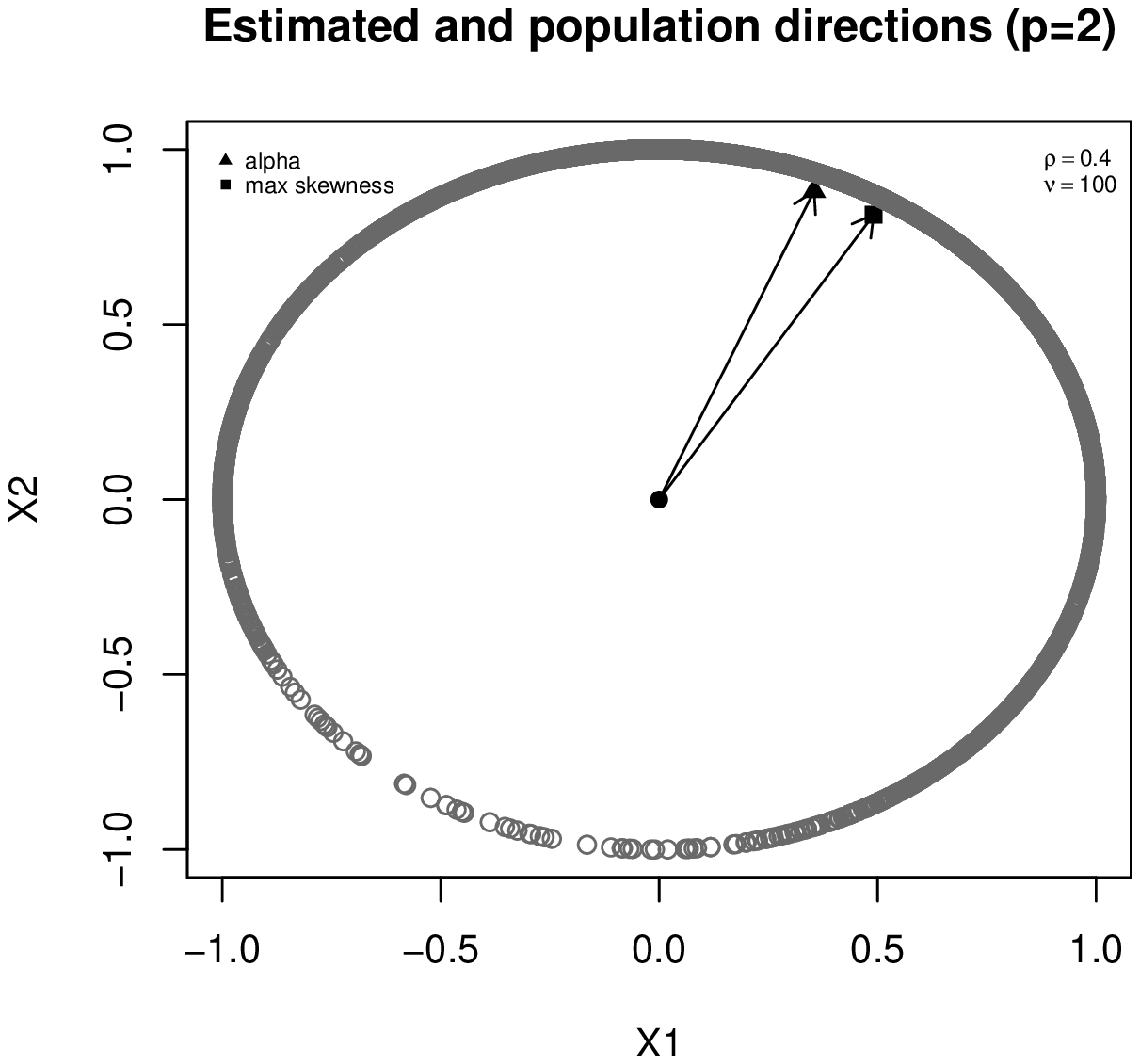}
        \end{subfigure}%
        \begin{subfigure}[b]{0.5\textwidth}
                \centering
                \includegraphics[height=4cm,width=4cm]{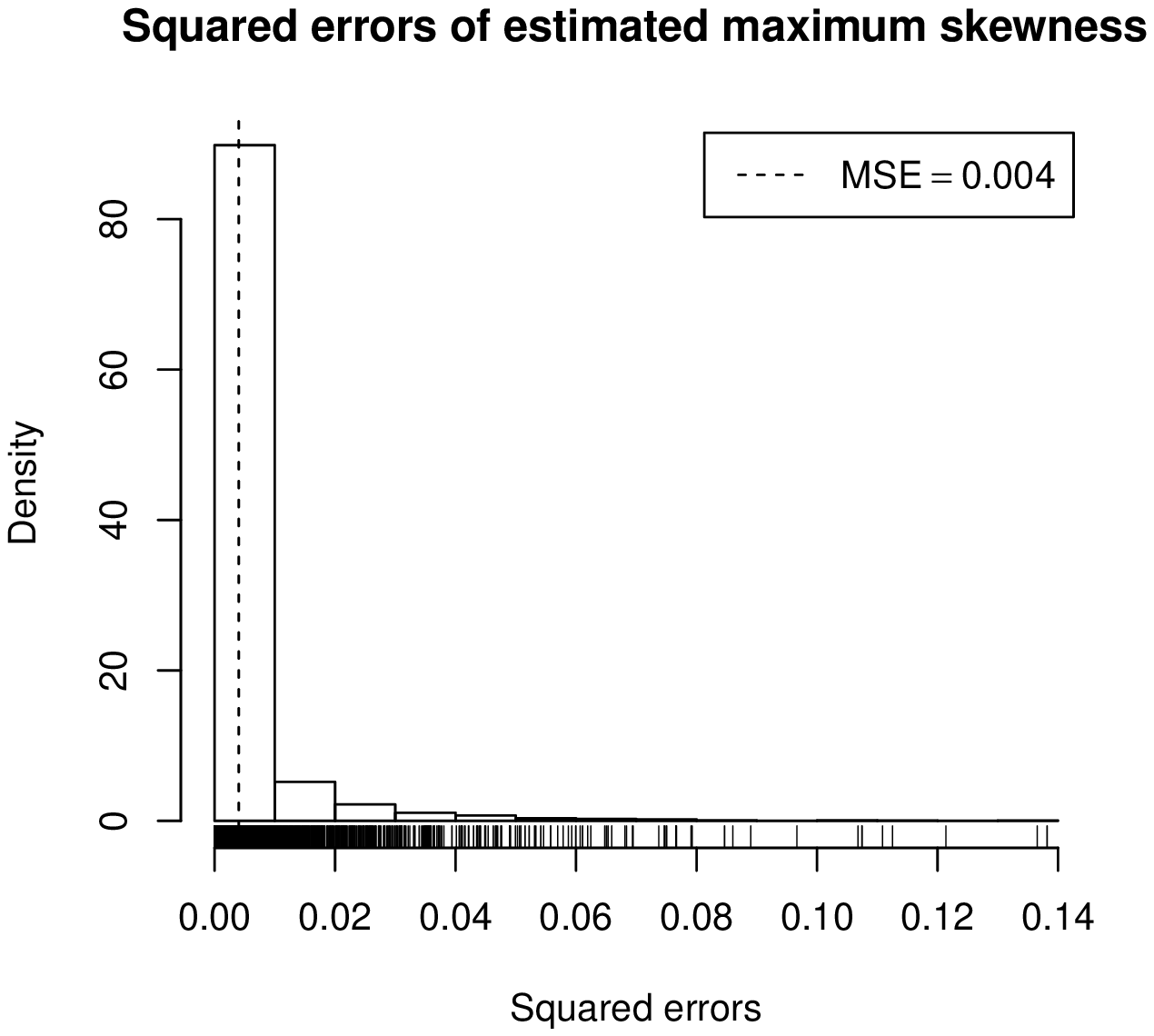}
        \end{subfigure}
        \begin{subfigure}[b]{0.5\textwidth}
                \centering
                \includegraphics[height=4cm,width=4cm]{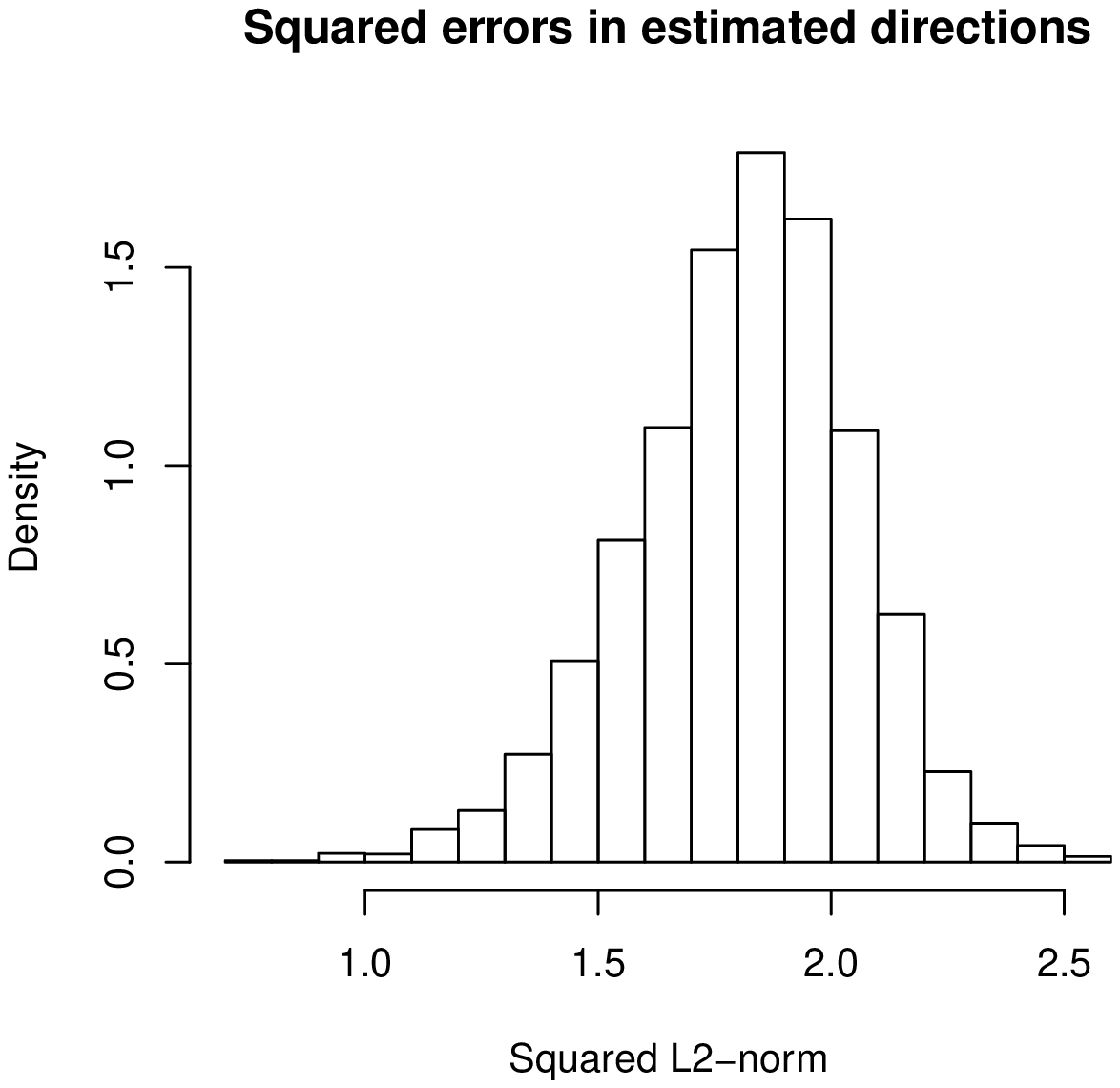}
        \end{subfigure}%
        \begin{subfigure}[b]{0.5\textwidth}
                \centering
                \includegraphics[height=4cm,width=4cm]{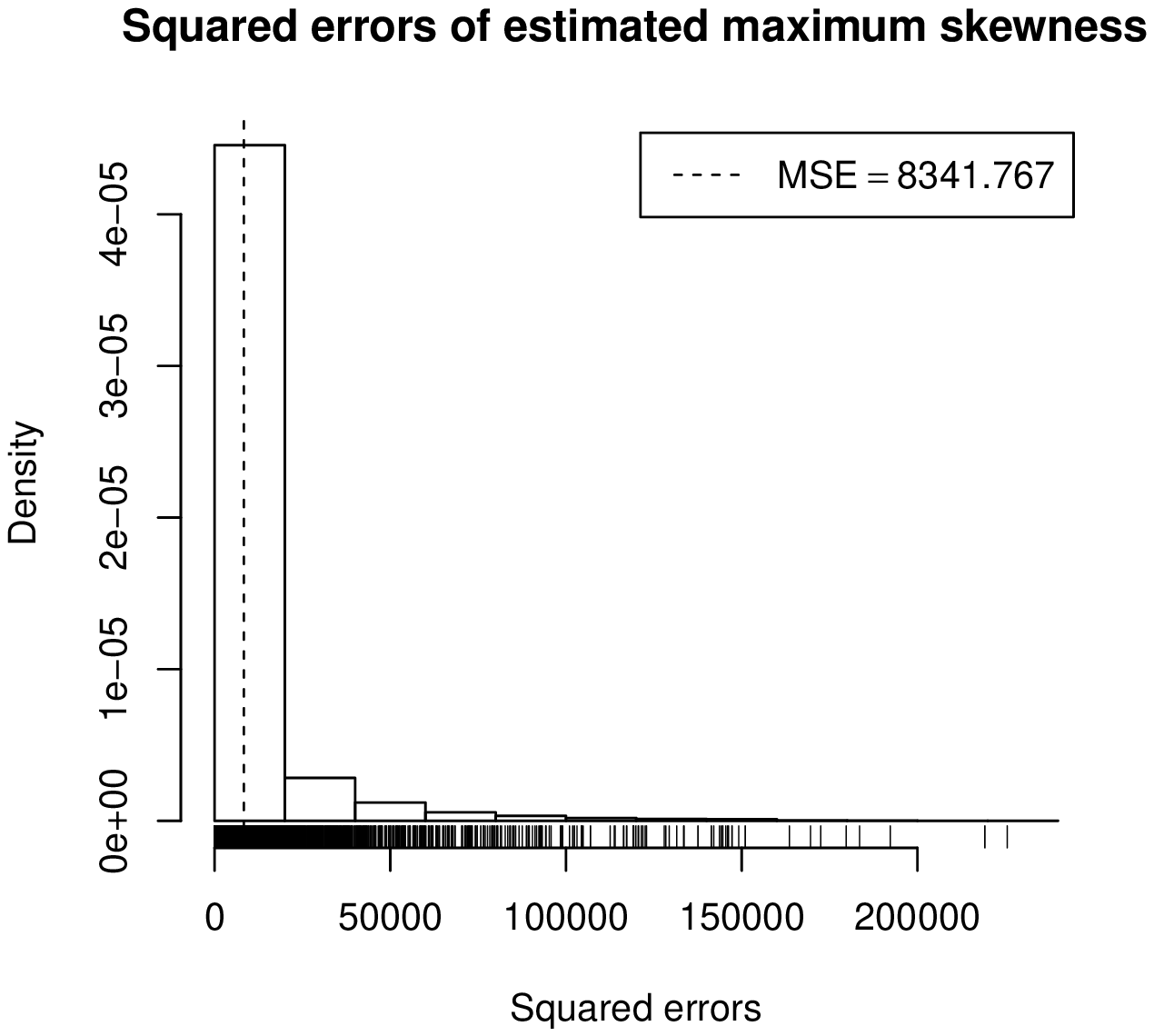}
        \end{subfigure}
        \\
        \begin{subfigure}[b]{0.5\textwidth}
                \centering
                \includegraphics[height=4cm,width=4cm]{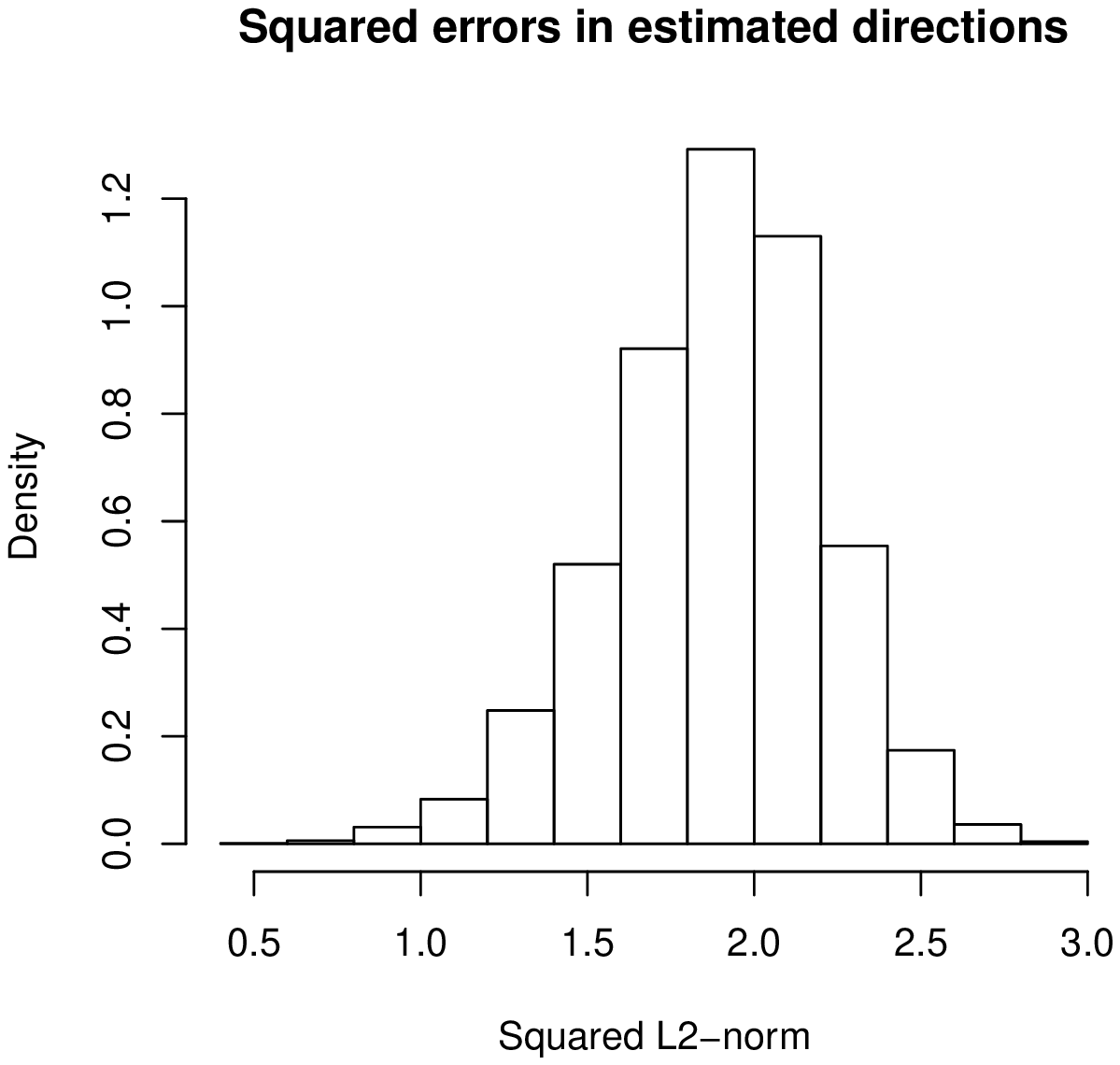}
        \end{subfigure}%
        \begin{subfigure}[b]{0.5\textwidth}
                \centering
                \includegraphics[height=4cm,width=4cm]{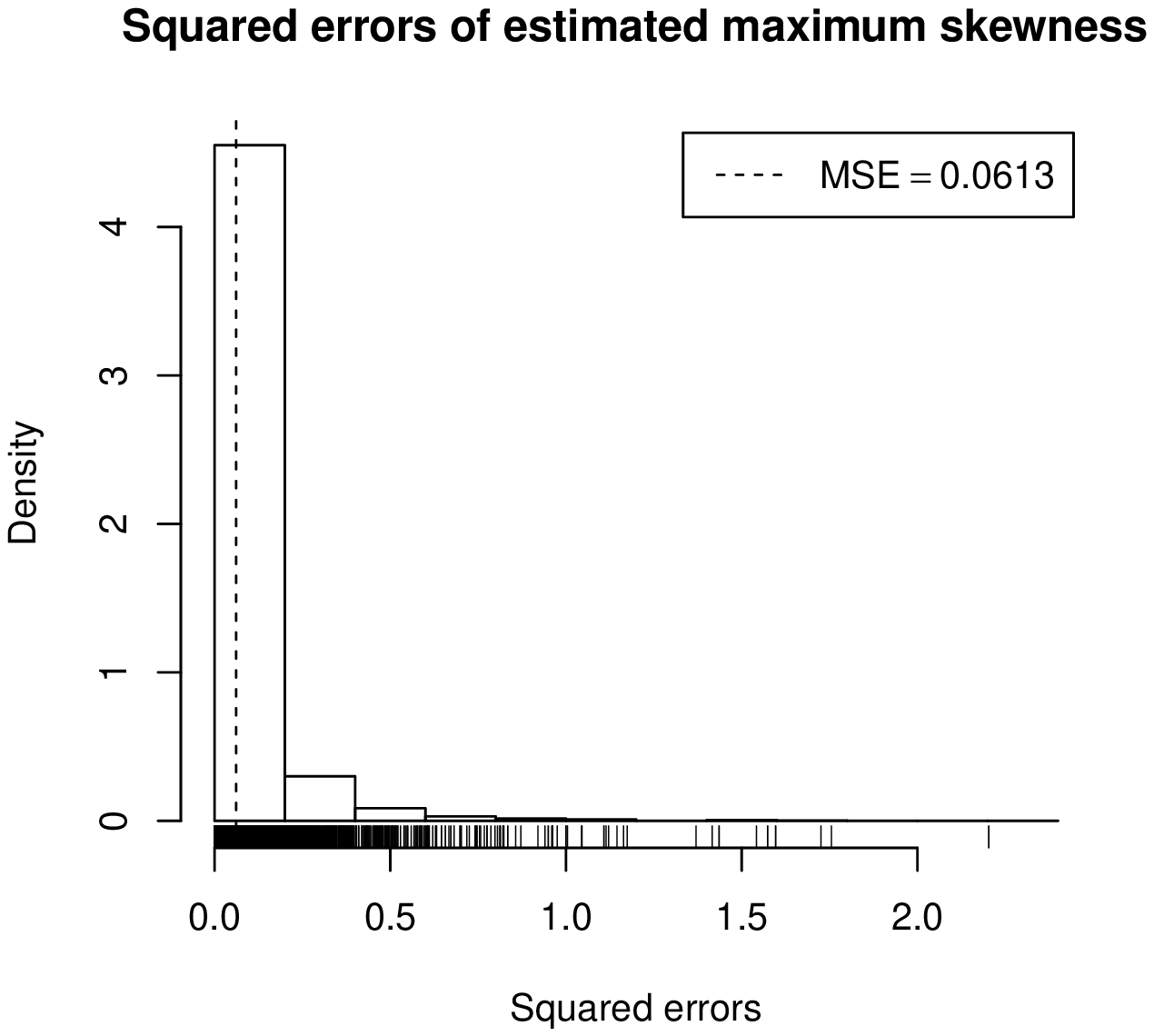}
        \end{subfigure}
        \caption{Plots when $n=500$  in Table \ref{TableRho040}: clock plots of the estimated directions and histograms of the squared error in the estimation of the maximal skewness when $p=2, \nu =4$ (first row) and $p=2, \nu =100$ (second row), as well as histograms of the squared $L2$-norm error of the estimated direction and the squared error of the maximal skewness when $p=18, \nu =4$ (third row) and $p=18, \nu =100$ (fourth row).}
        \label{PlotsRho040}
\end{figure}

\begin{figure}[h]
         \begin{subfigure}[b]{0.5\textwidth}
                \centering
                \includegraphics[height=4cm,width=4cm]{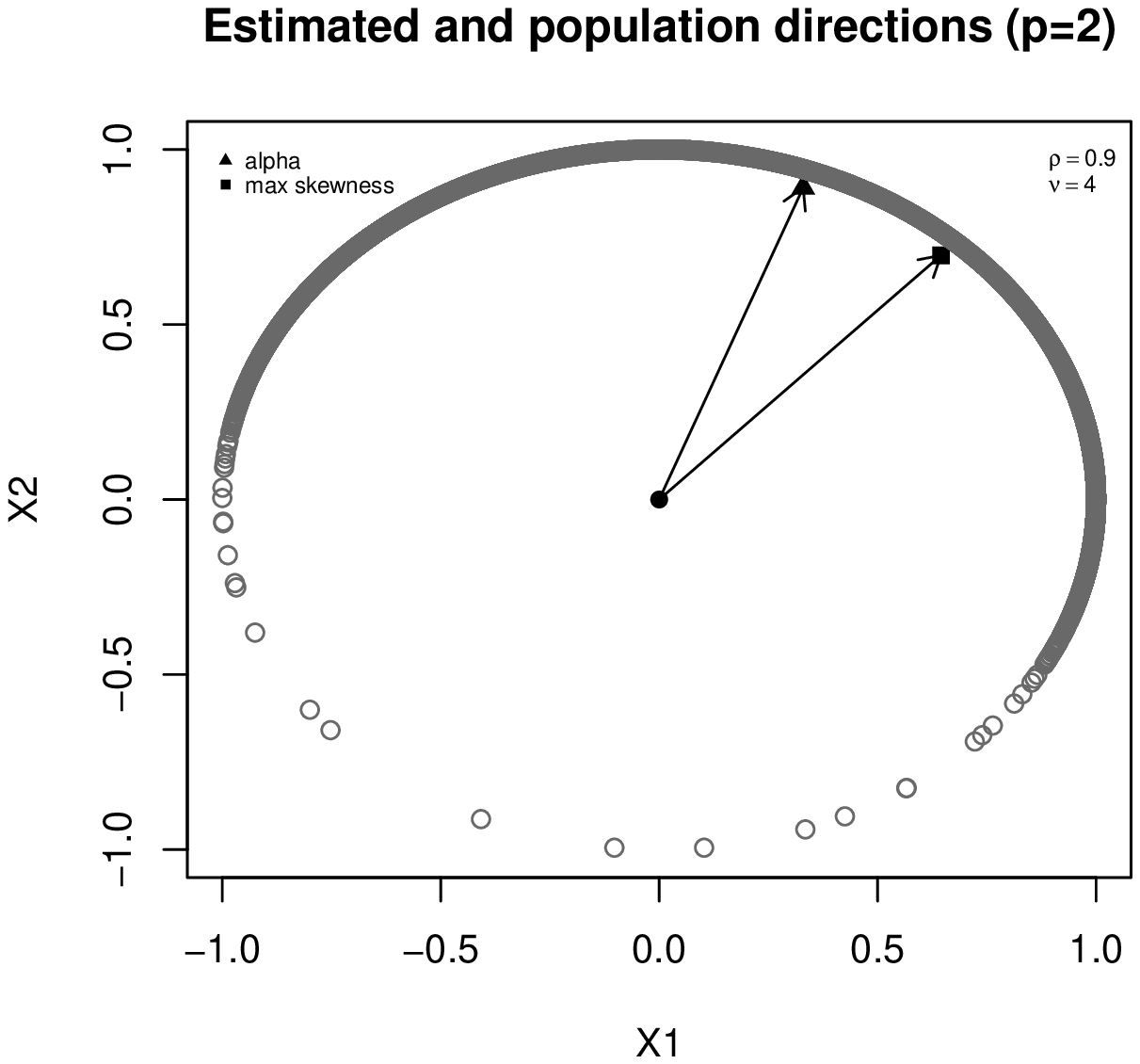}
        \end{subfigure}%
        \begin{subfigure}[b]{0.5\textwidth}
                \centering
                \includegraphics[height=4cm,width=4cm]{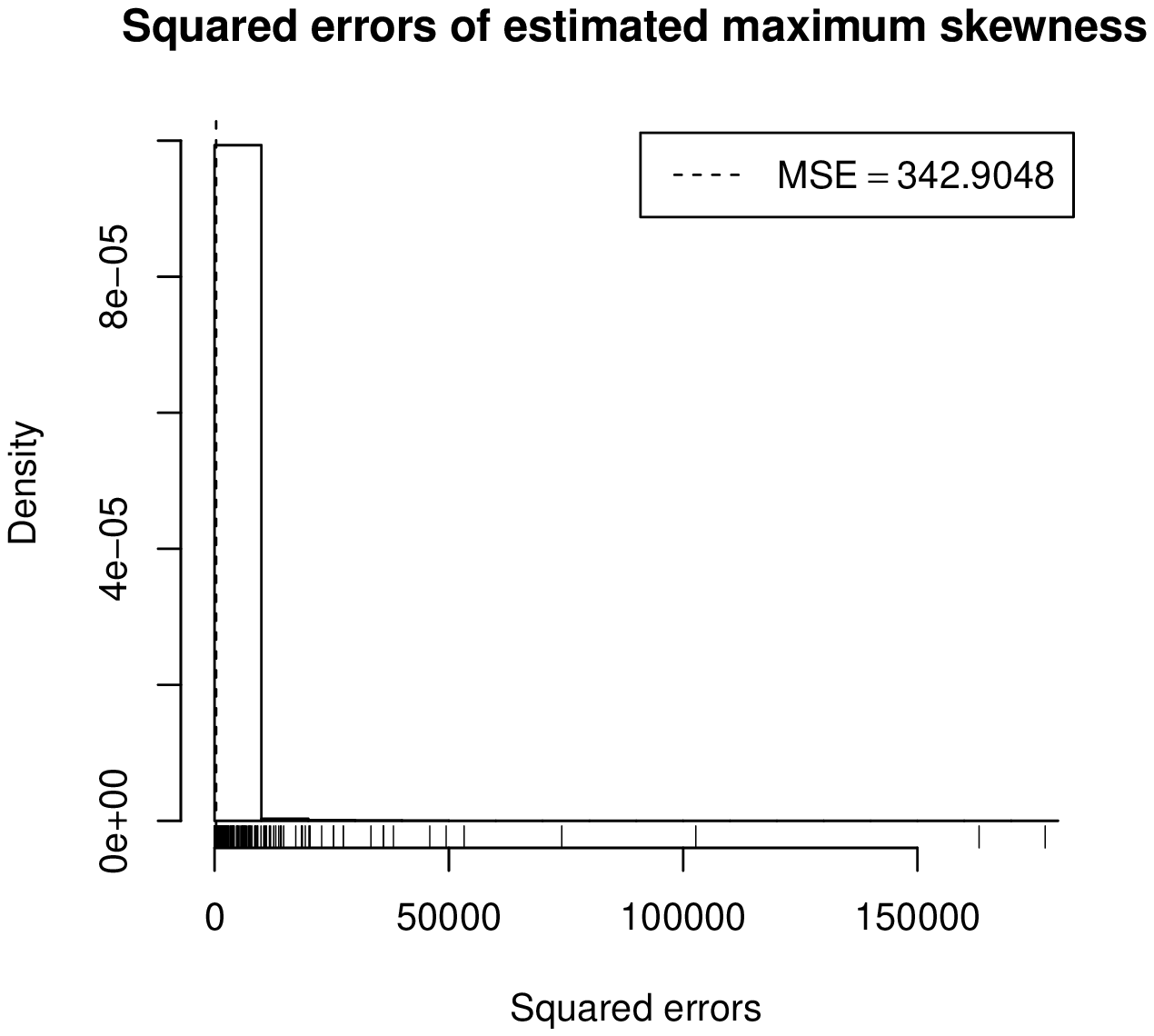}
        \end{subfigure}
        \\
        \begin{subfigure}[b]{0.5\textwidth}
                \centering
                \includegraphics[height=4cm,width=4cm]{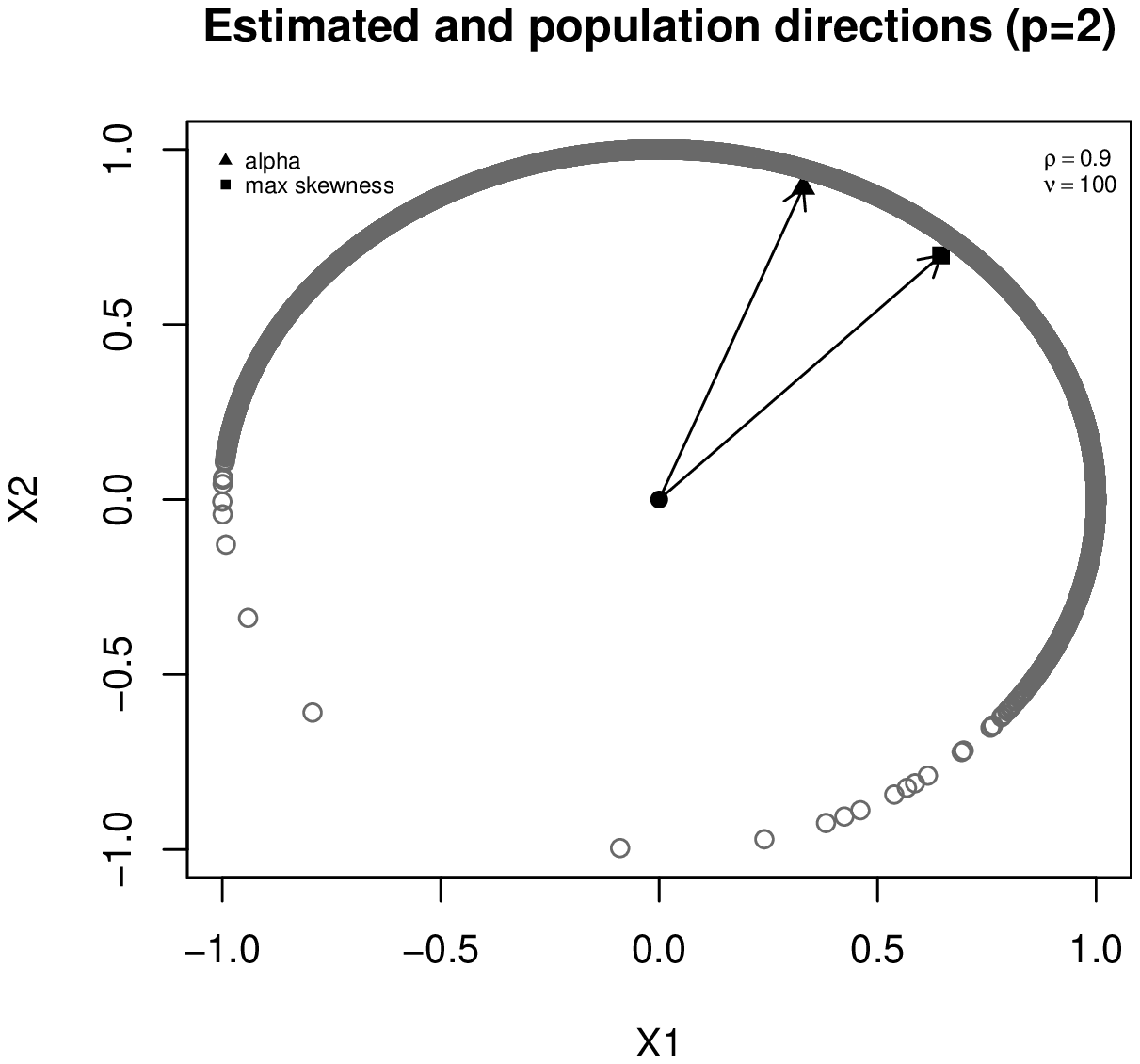}
        \end{subfigure}%
        \begin{subfigure}[b]{0.5\textwidth}
                \centering
                \includegraphics[height=4cm,width=4cm]{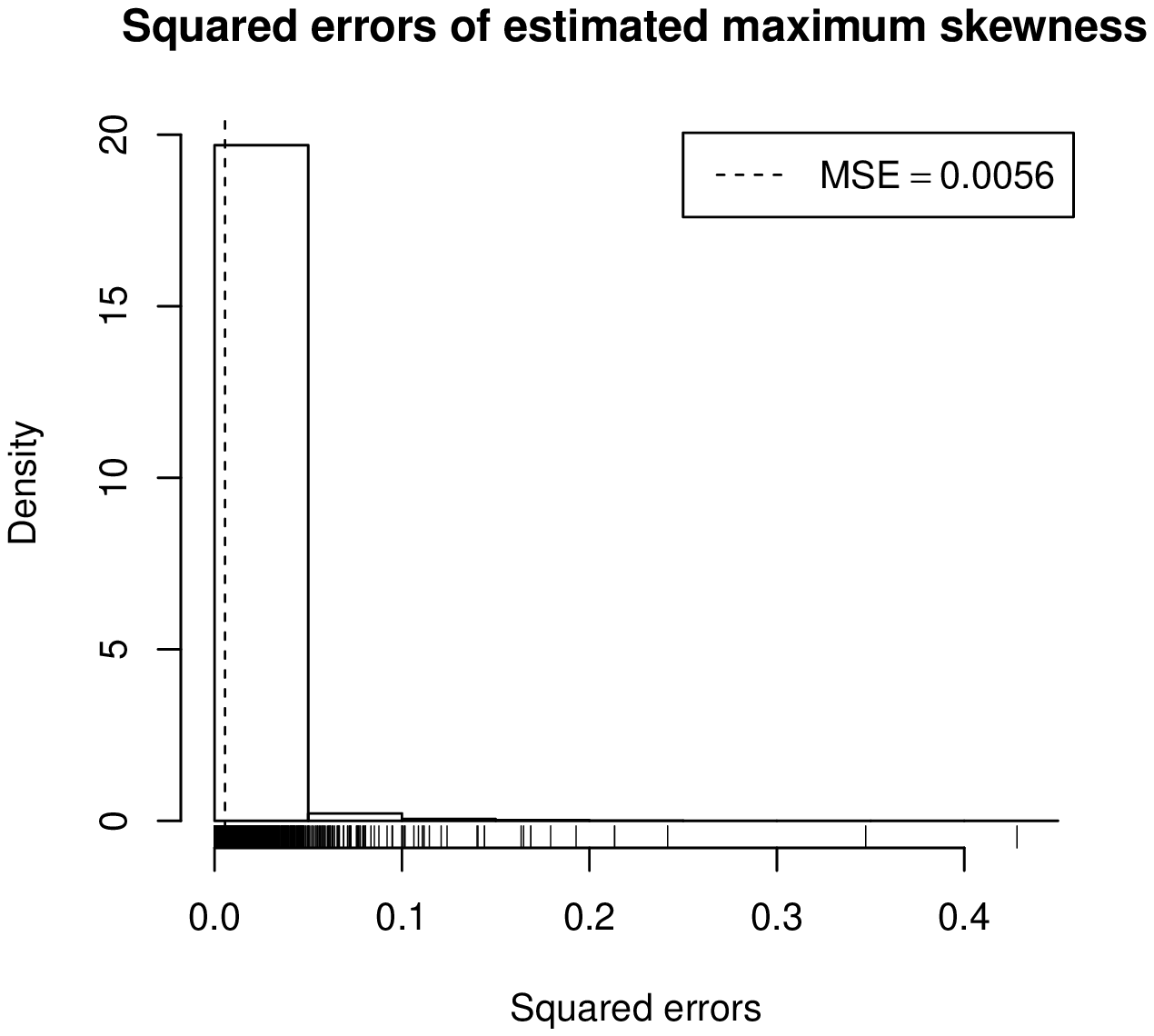}
        \end{subfigure}
        \begin{subfigure}[b]{0.5\textwidth}
                \centering
                \includegraphics[height=4cm,width=4cm]{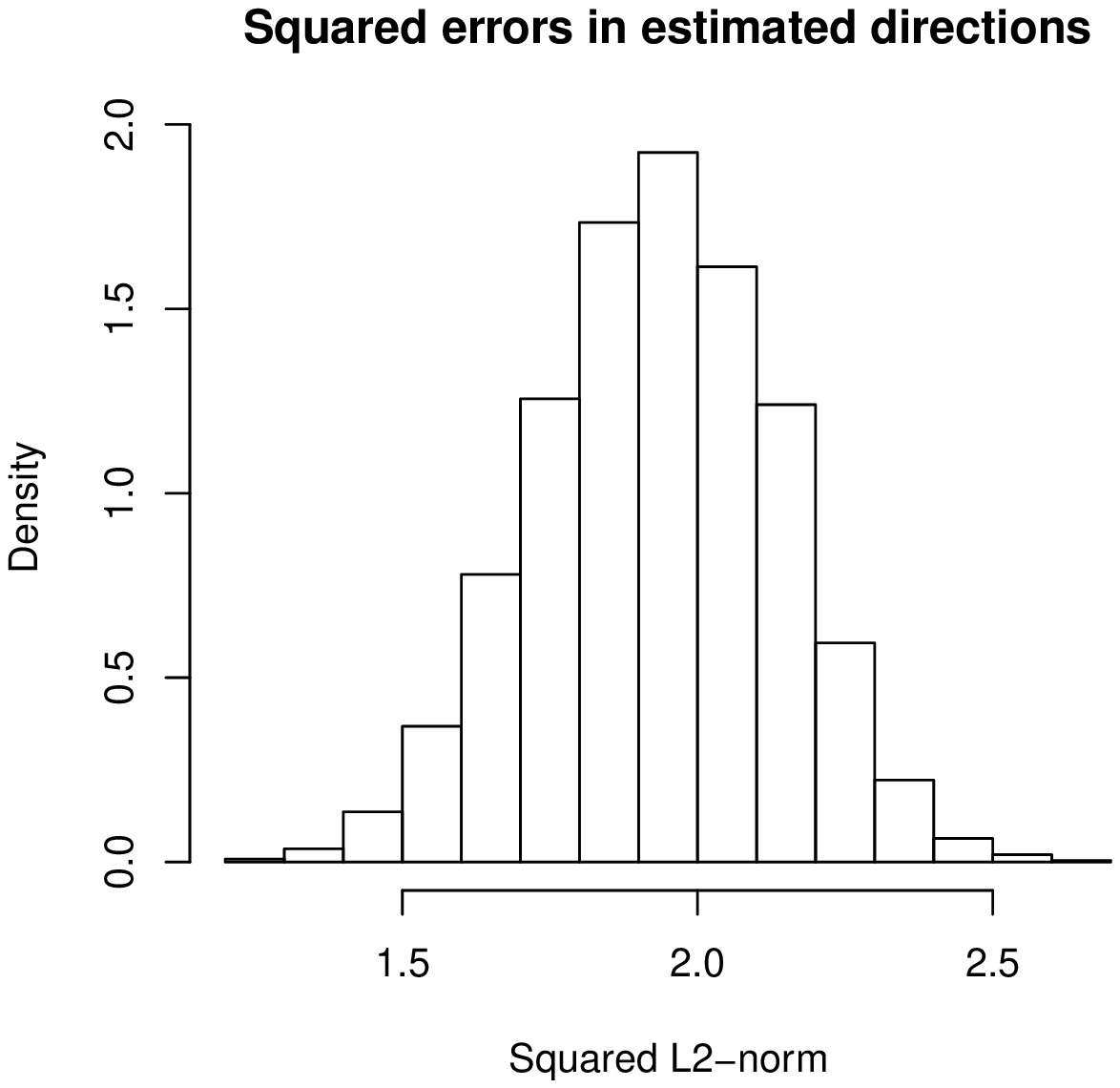}
        \end{subfigure}%
        \begin{subfigure}[b]{0.5\textwidth}
                \centering
                \includegraphics[height=4cm,width=4cm]{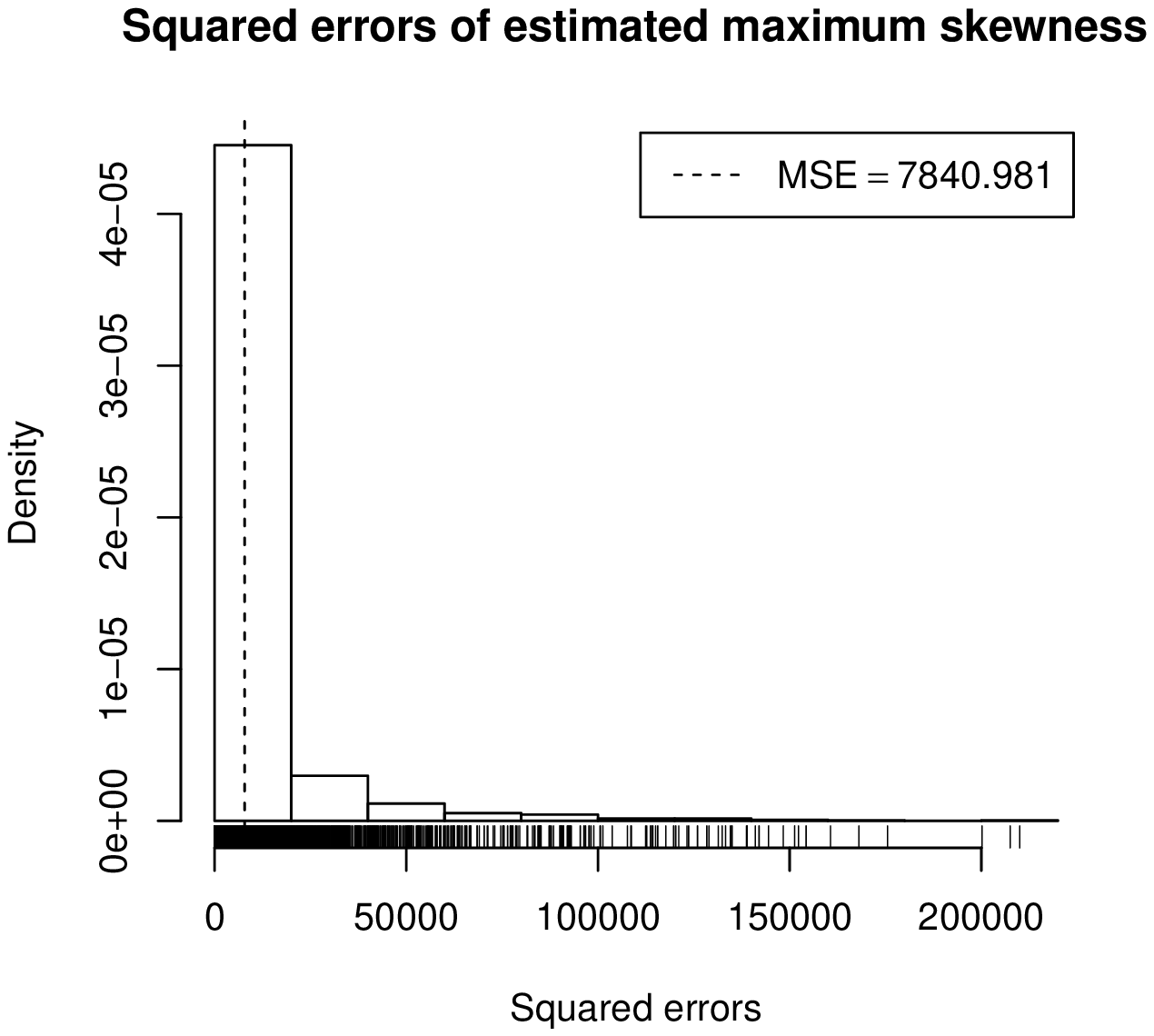}
        \end{subfigure}
        \\
        \begin{subfigure}[b]{0.5\textwidth}
                \centering
                \includegraphics[height=4cm,width=4cm]{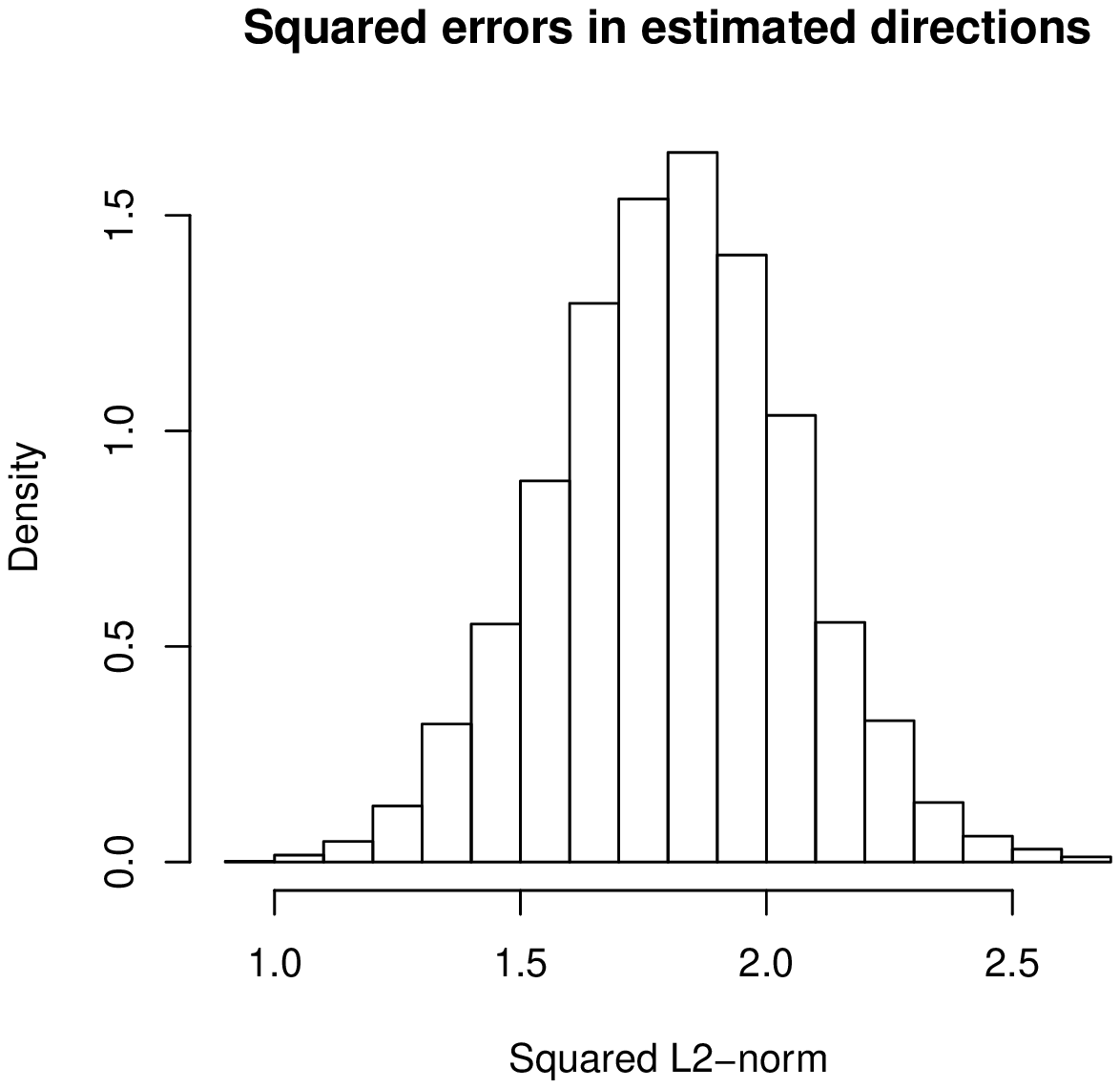}
        \end{subfigure}%
        \begin{subfigure}[b]{0.5\textwidth}
                \centering
                \includegraphics[height=4cm,width=4cm]{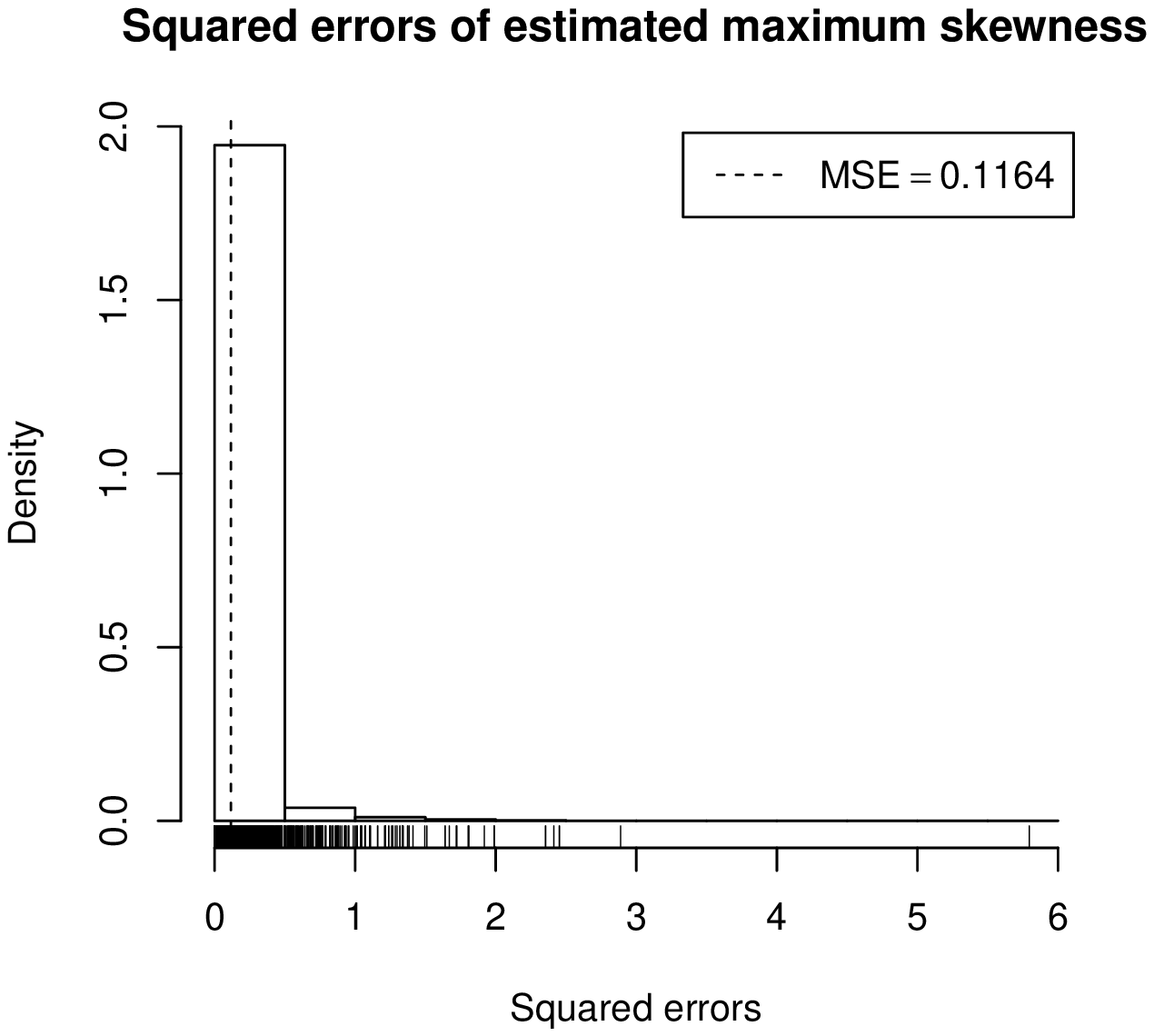}
        \end{subfigure}
        \caption{Plots when $n=500$  in Table \ref{TableRho090}: clock plots of the estimated directions and histograms of the squared error in the estimation of the maximal skewness when $p=2, \nu =4$ (first row) and $p=2, \nu =100$ (second row), as well as histograms of the squared $L2$-norm error of the estimated direction and the squared error of the maximal skewness when $p=18, \nu =4$ (third row) and $p=18, \nu =100$ (fourth row).}
        \label{PlotsRho090}
\end{figure}

\section{Summary and concluding remarks}
\label{conclusion}

In this paper we have addressed the problem of finding directions yielding the projection with maximum skewness for vectors that follow a multivariate SMSN distribution. A quite simple condition on the moments of the mixing variable is proposed; it ensures the main contribution of the paper which states that the maximal skewness direction is proportional to the shape vector $\boldsymbol\eta^\prime=\boldsymbol\alpha^\prime\boldsymbol\omega^{-1}$ that injects the directional asymmetry into the model. This is the case for some well-known multivariate distributions within the SMSN family, which include the skew normal, skew-t, skew double exponential and skew-slash distributions. The paper contributes to the field extending previous work for Skew Normal and Extended Skew Normal vectors \citep{Loperfido2010,FranceschiniLoperfido2014}, opening the road to move forward in the skewness based projection pursuit problem, both from the theoretical and inferential viewpoints, when the underlying multivariate model belongs to a wide rich and flexible class of distributions that account for the non normality of the data through tail weight and shape asymmetry parameters simultaneously.

\bibliographystyle{model1b-num-names}
\bibliography{SMSN_PP}

\begin{thebibliography}{20}
\expandafter\ifx\csname natexlab\endcsname\relax\def\natexlab#1{#1}\fi
\providecommand{\bibinfo}[2]{#2}
\ifx\xfnm\relax \def\xfnm[#1]{\unskip,\space#1}\fi
\bibitem[{Arevalillo and Navarro(2015)}]{ArevalilloNavarro2015}
\bibinfo{author}{J.M. Arevalillo}, \bibinfo{author}{H.~Navarro},
  \bibinfo{title}{A note on the direction maximizing skewness in multivariate
  skew-t vectors}, \bibinfo{journal}{Statistics \& Probability Letters}
  \bibinfo{volume}{96} (\bibinfo{year}{2015}) \bibinfo{pages}{328--332}.
\bibitem[{Azzalini(2005)}]{Azzalini2005}
\bibinfo{author}{A.~Azzalini}, \bibinfo{title}{The skew-normal distribution and
  related multivariate families}, \bibinfo{journal}{Scandinavian Journal of
  Statistics} \bibinfo{volume}{32} (\bibinfo{year}{2005})
  \bibinfo{pages}{159--188}.
\bibitem[{Azzalini and Capitanio(1999)}]{AzzaliniCapitanio1999}
\bibinfo{author}{A.~Azzalini}, \bibinfo{author}{A.~Capitanio},
  \bibinfo{title}{{Statistical applications of the multivariate skew normal
  distribution}}, \bibinfo{journal}{Journal of the Royal Statistical Society
  Series B} \bibinfo{volume}{61} (\bibinfo{year}{1999})
  \bibinfo{pages}{579--602}.
\bibitem[{Azzalini and Capitanio(2003)}]{AzzaliniCapitanio2003}
\bibinfo{author}{A.~Azzalini}, \bibinfo{author}{A.~Capitanio},
  \bibinfo{title}{Distributions generated by perturbation of symmetry with
  emphasis on a multivariate skew t-distribution}, \bibinfo{journal}{Journal of
  the Royal Statistical Society Series B} \bibinfo{volume}{65}
  (\bibinfo{year}{2003}) \bibinfo{pages}{367--389}.
\bibitem[{Azzalini and Capitanio(2014)}]{AzzaliniCapitanio2014}
\bibinfo{author}{A.~Azzalini}, \bibinfo{author}{A.~Capitanio},
  \bibinfo{title}{The Skew-Normal and Related Families},
  \bibinfo{publisher}{IMS monographs. Cambridge University Press},
  \bibinfo{year}{2014}.
\bibitem[{Azzalini and Dalla~Valle(1996)}]{AZZALINI1996}
\bibinfo{author}{A.~Azzalini}, \bibinfo{author}{A.~Dalla~Valle},
  \bibinfo{title}{The multivariate skew-normal distribution},
  \bibinfo{journal}{Biometrika} \bibinfo{volume}{83} (\bibinfo{year}{1996})
  \bibinfo{pages}{715--726}.
\bibitem[{Balakrishnan et~al.(2014)Balakrishnan, Capitanio and
  Scarpa}]{BalakrishnanCapitanioScarpa}
\bibinfo{author}{N.~Balakrishnan}, \bibinfo{author}{A.~Capitanio},
  \bibinfo{author}{B.~Scarpa}, \bibinfo{title}{A test for multivariate
  skew-normality based on its canonical form}, \bibinfo{journal}{Journal of
  Multivariate Analysis} \bibinfo{volume}{128} (\bibinfo{year}{2014})
  \bibinfo{pages}{19--32}.
\bibitem[{Balakrishnan and Scarpa(2012)}]{BalakrishnanScarpa}
\bibinfo{author}{N.~Balakrishnan}, \bibinfo{author}{B.~Scarpa},
  \bibinfo{title}{Multivariate measures of skewness for the skew-normal
  distribution}, \bibinfo{journal}{Journal of Multivariate Analysis}
  \bibinfo{volume}{104} (\bibinfo{year}{2012}) \bibinfo{pages}{73--87}.
\bibitem[{Branco and Dey(2001)}]{BrancoDey2001}
\bibinfo{author}{M.D. Branco}, \bibinfo{author}{D.K. Dey}, \bibinfo{title}{A
  general class of multivariate skew-elliptical distributions},
  \bibinfo{journal}{Journal of Multivariate Analysis} \bibinfo{volume}{79}
  (\bibinfo{year}{2001}) \bibinfo{pages}{99 -- 113}.
\bibitem[{Capitanio(2012)}]{Capitanio:arXiv1207.0797}
\bibinfo{author}{A.~Capitanio}, \bibinfo{title}{On the canonical form of scale
  mixtures of skew-normal distributions}, \bibinfo{journal}{arXiv/1207.0797}
  (\bibinfo{year}{2012}).
\bibitem[{Capitanio et~al.(2003)Capitanio, Azzalini and
  Stanghellini}]{CapitanioAzzaliniStanghellini}
\bibinfo{author}{A.~Capitanio}, \bibinfo{author}{A.~Azzalini},
  \bibinfo{author}{E.~Stanghellini}, \bibinfo{title}{Graphical models for
  skew-normal variates}, \bibinfo{journal}{Scandinavian Journal of Statistics}
  \bibinfo{volume}{30} (\bibinfo{year}{2003}) \bibinfo{pages}{129--144}.
\bibitem[{Contreras-Reyes and Arellano-Valle(2012)}]{Contreras2012}
\bibinfo{author}{J.E. Contreras-Reyes}, \bibinfo{author}{R.B. Arellano-Valle},
  \bibinfo{title}{Kullback-\uppercase{L}eibler divergence measure for
  multivariate skew-normal distributions}, \bibinfo{journal}{Entropy}
  \bibinfo{volume}{14} (\bibinfo{year}{2012}) \bibinfo{pages}{1606--1626}.
\bibitem[{Franceschini and Loperfido(2014)}]{FranceschiniLoperfido2014}
\bibinfo{author}{C.~Franceschini}, \bibinfo{author}{N.~Loperfido},
  \bibinfo{title}{Testing for normality when the sampled distribution is
  extended skew-normal}, \bibinfo{title}{Testing for Normality When the Sampled
  Distribution Is Extended Skew-Normal}, \bibinfo{publisher}{Springer
  International Publishing}, \bibinfo{year}{2014}, pp.
  \bibinfo{pages}{159--169}.
\bibitem[{Franceschini and Loperfido(2016)}]{MaxSkew}
\bibinfo{author}{C.~Franceschini}, \bibinfo{author}{N.~Loperfido},
  \bibinfo{title}{MaxSkew: Orthogonal Data Projections with Maximal Skewness},
  \bibinfo{year}{2016}. \bibinfo{note}{{R} package version 1.0}.
\bibitem[{G\'{o}mez-S\'{a}nchez-Manzano
  et~al.(2006)G\'{o}mez-S\'{a}nchez-Manzano, G\'{o}mez-Villegas and
  Mar\'{i}n}]{Gomez2006}
\bibinfo{author}{E.~G\'{o}mez-S\'{a}nchez-Manzano}, \bibinfo{author}{M.A.
  G\'{o}mez-Villegas}, \bibinfo{author}{J.M. Mar\'{i}n},
  \bibinfo{title}{Sequences of elliptical distributions and mixtures of normal
  distributions}, \bibinfo{journal}{Journal of Multivariate Analysis}
  \bibinfo{volume}{97} (\bibinfo{year}{2006}) \bibinfo{pages}{295--310}.
\bibitem[{Huber(1985)}]{Huber1985}
\bibinfo{author}{P.J. Huber}, \bibinfo{title}{Projection pursuit},
  \bibinfo{journal}{The Annals of Statistics} \bibinfo{volume}{13}
  (\bibinfo{year}{1985}) \bibinfo{pages}{435--475}.
\bibitem[{Loperfido(2010)}]{Loperfido2010}
\bibinfo{author}{N.~Loperfido}, \bibinfo{title}{Canonical transformations of
  skew-normal variates}, \bibinfo{journal}{TEST} \bibinfo{volume}{19}
  (\bibinfo{year}{2010}) \bibinfo{pages}{146--165}.
\bibitem[{Malkovich and Afifi(1973)}]{MalkovichAfifi1973}
\bibinfo{author}{J.F. Malkovich}, \bibinfo{author}{A.A. Afifi},
  \bibinfo{title}{On tests for multivariate normality},
  \bibinfo{journal}{Journal of the American Statistical Association}
  \bibinfo{volume}{68} (\bibinfo{year}{1973}) \bibinfo{pages}{176--179}.
\bibitem[{Merkle(1998)}]{Merkle1998}
\bibinfo{author}{M.~Merkle}, \bibinfo{title}{Conditions for convexity of a
  derivative and some applications to the gamma function},
  \bibinfo{journal}{aequationes mathematicae} \bibinfo{volume}{55}
  (\bibinfo{year}{1998}) \bibinfo{pages}{273--280}.
\bibitem[{Wang and Genton(2006)}]{WangGenton2006}
\bibinfo{author}{J.~Wang}, \bibinfo{author}{M.G. Genton}, \bibinfo{title}{The
  multivariate skew-slash distribution}, \bibinfo{journal}{Journal of
  Statistical Planning and Inference} \bibinfo{volume}{136}
  (\bibinfo{year}{2006}) \bibinfo{pages}{209--220}.

\end{thebibliography}







\end{document}